\documentclass[10pt,a4paper]{article}
\usepackage[english]{babel}
\usepackage{graphicx}
\usepackage{latexsym}
\newcommand{\alt}{\mathbin{\lower 3pt\hbox
   {$\rlap{\raise 5pt\hbox{$\char'074$}}\mathchar"7218$}}}
\newcommand{\agt}{\mathbin{\lower 3pt\hbox
   {$\rlap{\raise 5pt\hbox{$\char'076$}}\mathchar"7218$}}}

\begin{document}
\setcounter{footnote}{0}
\setcounter{equation}{0}
\setcounter{figure}{0}
\setcounter{table}{0}
\vspace*{5mm}

\begin{center}
{\large\bf A thorny path of field theory:\\
from triviality to interaction and confinement}

\vspace{4mm}
\vspace{4mm}
I. M. Suslov \\
Kapitza Institute for Physical Problems,
\\  Moscow, Russia \\
%\vspace{6mm}
% E-mail: suslov@kapitza.ras.ru
\vspace{1mm}
\end{center}

\begin{center}
\begin{minipage}{135mm}
{\bf Abstract } \\
Summation of the perturbation series for the
Gell-Mann--Low function $\beta(g)$ of $\phi^4$ theory
leads to the asymptotics $\beta(g)=\beta_\infty g^\alpha$ at
$g\to\infty$, where $\alpha\approx 1$ for space dimensions
$d=2,3,4$.  The natural hypothesis arises, that asymptotic
behavior is $\beta(g) \sim g$ for all $d$.  Consideration of the
"toy" zero-dimensional model confirms the hypothesis and reveals
the origin of this result: it is related with a zero of a certain
functional integral.  This mechanism remains valid for arbitrary
space dimensionality $d$. The same result for the asymptotics is
obtained for  explicitly accepted lattice regularization, while
the use of high-temperature expansions allows to calculate the
whole $\beta$-function. As a result, the $\beta$-function  of
four-dimensional $\phi^4$ theory is appeared to be
non-alternating and has a linear asymptotics at infinity.
The analogous situation is valid for QED.
According to the Bogoliubov and Shirkov classification, it means
possibility to construct the continuous theory with finite
interaction at large distances.  This conclusion is in visible
contradiction with  the lattice results indicating  triviality of
$\phi^4$ theory.  This contradiction is resolved by a special
character of renormalizability in $\phi^4$ theory:  to obtain the
continuous renormalized theory, there is no need to eliminate a
lattice from the bare theory.  In fact, such kind of
renormalizability is not accidental  and can be understood in the
framework of Wilson's many-parameter renormalization group.
Application of these ideas to QCD shows  that  Wilson's theory
of confinement is not purely illustrative,
but has a direct relation to a real situation.  As a result, the
problem of  analytical proof of  confinement and a mass gap
can be considered as solved, at least on the physical level of
rigor.

 \end{minipage} \end{center} %\vspace{5mm}

%\vspace*{1.5mm}
%PACS 11.10Kk, 11.15.Pg, 11.15.Me, 64.60.Fr, 75.10.Hk
%\vspace*{1.5mm}

\newpage
\begin{center}
{\bf 1. Introduction}
\end{center}

In 1954  Landau, Abrikosov and Khalatnikov \cite{101} derived
the famous relation between the bare charge $g_0$ and
observable charge $g$ for renormalizable field theories:
$$
g=\frac{g_0}{1+\beta_2 g_0 \ln \Lambda/m}  \,,
\eqno(1)
$$
where $m$ is the mass of the particle, and $\Lambda$ is
the momentum cut-off. The constant $\beta_2$ is positive in
$\phi^4$ theory and QED, so $g$ tends to zero in the limit
$\Lambda\to \infty$ for any finite $g_0$, i.e. the "zero
charge" situation takes place. In fact, the proper
interpretation of Eq.\,1 was given in \cite{101} and consists in
its inverting, so that $g_0$ is attributed to the length scale
$\Lambda^{-1}$ and is chosen to give a correct value of $g$:
$$
g_0=\frac{g}{1-\beta_2 g \ln \Lambda/m} \,.
\eqno(2)
$$
The growth of $g_0$ with  $\Lambda$ invalidates Eq.\,1
(obtained perturbatively) in the region $g_0\sim 1$,
and existence of "the Landau pole" in Eq.\,2 has no physical
sense.

A little later,  Landau and Pomeranchuk \cite{102}
put forward arguments on validity of Eq.\,1 for arbitrary
$g_0$. They have noticed that the constant limit for the
observable charge $g$ can be obtained in the limit
$g_0\to \infty$ from the functional integrals of the
lattice $\phi^4$ theory, if the quadratic in $\phi$ terms
are omitted in the action \cite{102}\,\footnote{\,In fact,
the limit $g_0\to \infty$ is
accompanied by the limit $m_0^2\to -\infty$ for the bare
mass, so the terms $m_0^2 \phi^2$ and $g_0 \phi^4$ are
equally significant  (Sec.\,6). Accepting different laws of growth for
$-m_0^2$, one can obtain different possibilities. On the
other hand, the strong coupling limit for $g$ can be attained for
complex $g_0$ (Sec.\,4) where arguments by Landau and Pomeranchuk
are not valid in principle. }.
On the other hand, the constant limit
$1/(\beta_2 \,\ln \Lambda/m)$ is reached with growth of $g_0$
already in the weak coupling region, which is described by Eq.\,1.
It looks that neglecting of quadratic terms is possible
already for $g_0\ll 1$, and it is all the more possible
for $g_0\agt 1$:  it gives a reason to consider Eq.\,1 to be
valid for arbitrary $g_0$. Analogous arguments are possible
in the case of QED \cite{102}.  These results lead Landau to
conclusion on fundamental deficiency of the field theoretical
description  \cite{103}.

This conclusion was questioned by Bogoliubov and Shirkov
\cite{104}, who noted that actual behavior of the charge
$g(L)$ as a function of the length  scale $L$ is determined by
the Gell-Mann -- Low equation
$$
-\frac{dg}{d \ln L} =\beta(g)=\beta_2 g^2+\beta_3 g^3+\ldots
\eqno(3)
$$
and depends on appearance of the function  $\beta(g)$. According
to the Bogoliubov and Shirkov classification \cite{104},
there are three qualitatively different possibilities (Fig.\,1):
\begin{figure}
\centerline{\includegraphics[width=5.5 in]{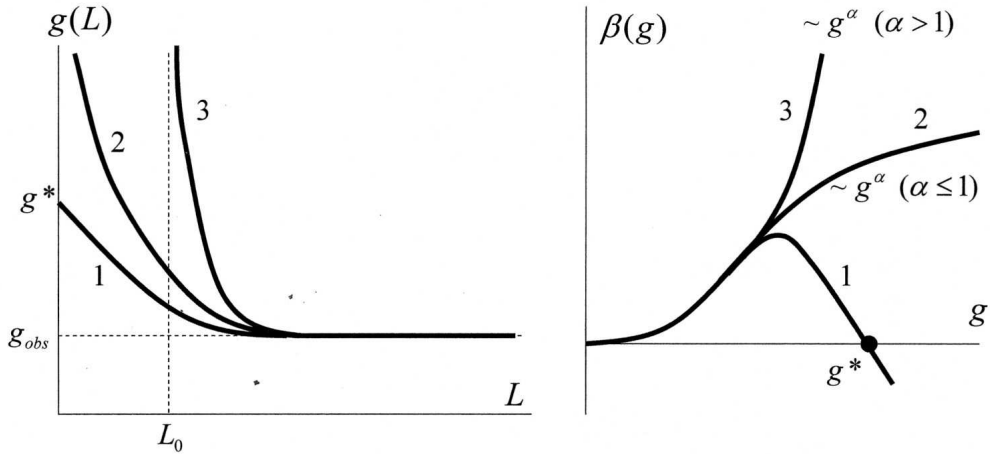}}
\caption{Three qualitatively different situations according to the
Bogoliubov and Shirkov classification. } \label{fig1}
\end{figure}
(i) if $\beta(g)$ has a  zero at some
point $g^*$, then the effective coupling  $g$ tends to
$ g^*$ at small $L$; (ii) if $\beta(g)$ is
non-alternating and has asymptotic behavior  $g^\alpha$
with $\alpha\le 1$, then $g(L)$ grows to
infinity; (iii) if non-alternating $\beta(g)$ behaves at infinity
as $ g^\alpha$ with $\alpha>1$, then $g(L) $ is divergent  at
some finite $L_0$ (the real Landau pole arises)
and  dependence  $g(L)$ is not defined at smaller distances:
the theory is internally inconsistent and a finite interaction
at large distances is impossible in the continual limit.
The latter case corresponds to the "zero charge" situation
in full theory, beyond its perturbative context. Realization
of this situation cannot be proved, because a behavior of the
$\beta$-function is unknown.\,\footnote{\,Equation (1) follows
from (3), if only the first term is retained in the right hand
side. It cannot be exact due to finiteness of $\beta_3$.}

In current literature these problems are discussed in relation
with the concept of "triviality", introduced by Wilson
\cite{105}. In the theory of critical phenomena, a finite
interaction is accepted at small length scales
corresponding to the lattice spacing, while Eq.\,3
is integrated in direction of large $L$. If  $\beta(g)$
is positive beyond the origin, then $g\to 0$ at large
distances and $\phi^4$ theory reduces to the trivial
Gaussian model: it corresponds to the absence of interaction
between large-scale fluctuations of the order parameter.
According to Wilson's renormalization group \cite{105a} such
triviality takes place for Euclidean $\phi^4$ theory in space
dimensions $d\ge 4$. Success of Wilson's $\epsilon$--expansion
\cite{105a} is
directly related with this triviality: for $d=4-\epsilon$,
interaction between large-scale fluctuations becomes finite but
small for $\epsilon\ll 1$.

In the weak coupling region, the $\beta$-function of
four-dimensional $\phi^4$ theory is positive and triviality
surely exists. In subsequent papers, Wilson set problem more
deeply:  does triviality for $d=4$ exist only for small $g_0$, or
has the global character?  Using logic of proof by contradiction,
he assumed existence of the boundary $g_f$ for the domain of
attraction of the Gaussian fixed point $g=0$ (which is equivalent
to alternating behavior for $\beta(g)$) and derived the
consequences convenient for numerical verification.  According to
his results \cite{105}, there are no indications on existence of
$g_f$.  Historically, it was the first real attempt to
investigate the strong coupling regime for $\phi^4$ theory and
the first evidence of non-alternating behavior of $\beta(g)$.

Another definition of triviality was given in the mathematical
papers  \cite{106}--\cite{108}. It corresponds to  true
triviality, i.e. impossibility in principle to construct
continuous theory with finite interaction at large distances. It
is equivalent to internal inconsistency in the Bogoliubov and
Shirkov sense, or Landau's "zero charge". It was rigorously
proved in \cite{106}--\cite{108} that $\phi^4$ theory is trivial
for  $d>4$ and nontrivial for $d<4$; using experience of these
proofs, some plausible arguments were given in favor of
triviality for $d=4$. From the physical point of view, the former
results are rather evident \cite{113}: triviality for $d>4$
follows from
nonrenormalizability of $\phi^4$ theory, while nontriviality  for
$d<4$ is a consequence of the nonzero root of $\beta(g)$,
whose existence is easily established for $d=4-\epsilon$ with
$\epsilon\ll 1$. These results  do not require any study of the
strong coupling region, and hence no propositions can be made for
the case $d=4$, where such investigation is obligatory.

It should be clear that two definitions of triviality are not
equivalent. Wilson triviality needs only
positiveness of the $\beta$--function for $g\ne 0$, while
true triviality demands in addition
the corresponding asymptotic behavior.  Wilson triviality
can be considered as firmly established (see a review and numerous
references in \cite{113}), while evidence of true triviality is not
extensive and allows different interpretation.
The most interesting example is shown in Fig.2: it gives
dependencies of the renormalized charge against the
bare one for fixed $\Lambda/m$  \cite{137} and looks as numerical
confirmation of argumentation by Landau and Pomeranchuk ($N$ is
proportional to $\Lambda/m$).
\begin{figure}
\centerline{\includegraphics[width=2.5 in]{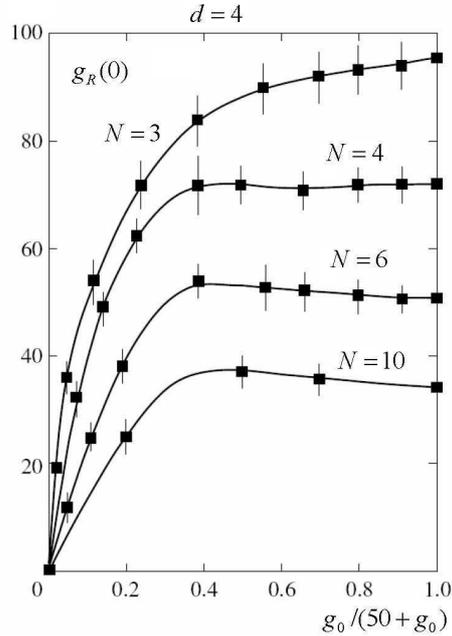}}
\caption{The renormalized charge $g_R(0)$ (estimated for zero
momenta) against the bare charge $g_0$ (corresponding to
interatomic spacing $a$) in four-dimensional $\phi^4$ theory
for fixed values of $Na$ and $m$ but different number
$N^4$ of lattice sites
(according to \cite{137}). } \label{fig2}
\end{figure}
More close inspection reveals that all results for finite
$g_0$ correspond to the parabolic portion of the
$\beta$-function\,\footnote{\,In $\phi^4$ theory, the "natural"
normalization of charge corresponds to the interaction term
written as $(16\pi^2/4!)g\phi^4$. In this case, the nearest
singularity in the Borel plane (Sec.\,2) lies at the unit distance
from the origin, and $\beta(g)$ is expected to change on the
scale of the order of unity. In fact, even in the natural normalization
the one-loop behavior appears to be somewhat dragged-out, and
approximately  quadratic dependence of $\beta(g)$ continues till
$g\sim 10$  (see Fig.4,$a$ below).
If the interaction term is  written as $g\phi^4/8$ or
$g\phi^4/4!$,  the boundary between "weak coupling" and
"strong coupling" regions lies at $g\sim 10^3$ instead of  $g\sim 1$.}
and do not manifest  essential deviations from Eq.\,1, while
the points for $g_0=\infty$ were obtained by reducing to the
Ising model, which is an ambiguous procedure.

In fact, two definitions of triviality were hopelessly mixed in
the literature (see Sec.\,8). As a result, to the end of 20-th
century a conviction in triviality of $\phi^4$ theory and QED
became predominated in literature. Below we overview the
comparatively new results \cite{109}--\cite{118} obtained after
2000, which prove the absence of true triviality for these theories.

It is clear from preceding discussion  that solution of the "zero
charge" problem needs calculation of the Gell-Mann -- Low function
$\beta(g)$ at arbitrary $g$, and in particular its asymptotic
behavior for $g\to\infty$. Approaches to this problem are
discussed in the next sections. Summation of the perturbation
series for $\beta(g)$ with the use of the Lipatov asymptotics
gives the positive $\beta$-function in four-dimensional $\phi^4$
theory and its asymptotic behavior $g^\alpha$ with $\alpha\approx
1$ (Sec.\,2).  The same result for $\alpha$ is obtained in
dimensions $d=2$ and $d=3$. The arising hypothesis $\beta(g)\sim
g$ for the asymptotic behavior is confirmed in the
zero-dimensional case (Sec.\,3) and extended to arbitrary
dimensions (Sec.\,4). The same approach allows to obtain the
$\beta$-function in QED (Sec.\,5). The problem of
complex-valuedness of the bare coupling constant is discussed in
Sec.\,6 and a scheme without complex parameters is formulated.
The latter involves the explicit lattice regularization and
reveals the surprising property in renormalizability of $\phi^4$
theory: the continual limit in the renormalized theory does not
demand the continual limit in the bare theory. This property
allows to give a final solution of the triviality problem
(Sec.\,8) and makes it possible to use  high-temperature
expansions for calculation of the $\beta$-function with good
precision (Sec.\,7). The character of renormalizability discovered
for $\phi^4$ theory is shown to have a general character
(Sec.\,9), which can be applied to justification of the Wilson
theory of confinement (Sec.\,10).

%\vspace{6mm}
\begin{center}
{\bf 2. Summation of perturbation series}
\end{center}

Let consider the typical problem in field theory applications.
A certain quantity $W(g)$ is given by its formal perturbation
expansion
$$
W(g)= \sum\limits_{N=0}^{\infty} \,W_N (-g)^N
\eqno(4)
$$
in powers of the coupling constant $g$. The coefficients
$W_N$ are given numerically and have the factorial asymptotics
at $N\to\infty$,
$$
W_N^{as}=c_0 a_0^N \Gamma(N+b_0) \,,
\eqno(5)
$$
which is a typical result obtained by the Lipatov method
\cite{1}. We want to find $W(g)$ for arbitrary $g$
while the radius of convergence for (4) is zero.

The standard summation procedure is based on the Borel
transformation: each term is divided and multiplied by $N!$,
the factorial in the numerator is replaced by the definition
of the gamma-function, then summation and integration are
interchanged,
$$
W(g)=\sum\limits_{N=0}^{\infty} W_N g^N =
     \sum\limits_{N=0}^{\infty} \frac{W_N}{N!}
     \int\limits_{0}^{\infty} dx \, x^N {\rm e}^{-x} g^N =
      \int\limits_{0}^{\infty} dx \,{\rm e}^{-x}
     \sum\limits_{N=0}^{\infty} \frac{W_N}{N!} (gx)^N \quad,
$$
and we have a series with a factorially improved convergence.
We can use $\Gamma(N+b)$ with arbitrary $b$ instead $N!$
and obtain the general Borel--Leroy transformation:
%in the Borel--Leroy form:
$$
W(g)=\int\limits_{0}^{\infty} dx e^{-x} x^{b-1} B(gx)\,,\qquad
$$
$$
B(z)=\sum\limits_{N=0}^{\infty} B_N (-z)^N\,,\qquad
B_N=\frac{W_N}{\Gamma(N+b)}\,.
\eqno(6)
$$
The function $W(g)$ is related with its Borel transform $B(z)$
by some integral transformation, while $B(z)$ is given by
a series with a factorially improved convergence.

It is easy to show that the Borel transform $B(z)$ has a
singularity at the point $z=-1/a_0$ (Fig.\,3,\,$a$)
determined by the parameter
$a_0$ in the Lipatov asymptotics (5). The series for $B(z)$
is convergent in the disk  $|z|<1/a_0$, while we should know it
on the positive semi-axis, in order to perform integration
in the Borel integral  (6); so we need  analytical
continuation of $B(z)$. Such analytical continuation is easy
if the coefficients $W_N$ are defined by some simple formula,
but it is a problem when they are given numerically.
\begin{figure}
\centerline{\includegraphics[width=3.5 in]{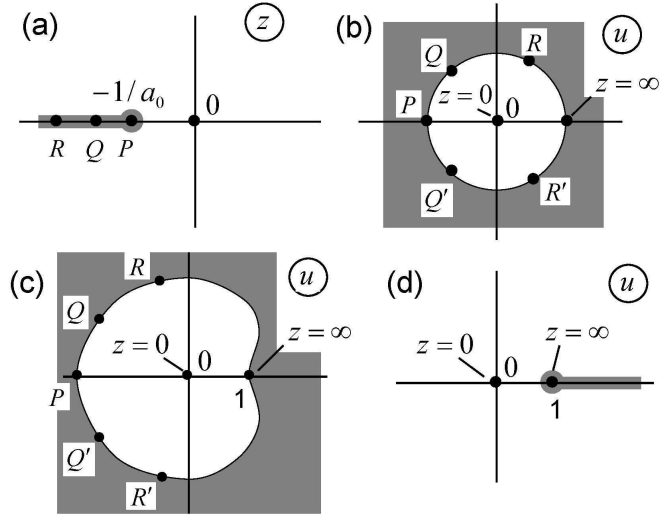}} \caption{
($a$) The Borel transform $B(z)$ is analytical in the complex
plane with the cut $(-\infty, -1/a_0)$; ($b$) Its domain of
analyticity can be conformally mapped to a unit disk in the
$u$ plane; ($c$) If analytic continuation is restricted to the
positive semi-axis, then a conformal mapping can be made to any
domain, for which the point $u =1$ is the nearest to the origin of
all boundary points; ($d$) An extreme case of such domain is the
$u$ plane with the cut $(1,\infty)$. }
\label{fig3} \end{figure}

The elegant solution of this problem was given by Le Guillou and
Zinn-Justin in 1977 \cite{2}. It is based on the hypothesis that
in field theory applications all singularities of $B(z)$ lie on
the negative semi-axis.
%, on the left from the point $z=-1/a$.
This hypothesis can be proved in the case of $\phi^4$ theory
\cite{3}.\,\footnote{\,Validity of this hypothesis
is frequently questioned in relation with possible
existence of renormalon singularities  \cite{200}.
Such  singularities can be easily obtained by summing some special
sequences of  diagrams, but their existence was never
proved, if all diagrams are taken into account \cite{200a}.
The given below results for the asymptotics of
$\beta(g)$ (Secs.\,4,\,5) are in agreement with a
general criterion for absence of renormalon singularities
\cite{202} and a proof of their absence  for $\phi^4$ theory
\cite{3} (see a detailed discussion in \cite{112}).} If such
analytical properties are accepted, we can make a conformal
transformation $z=f(u)$, mapping the complex plane with the cut
(Fig.\,3,\,$a$) into the unit disk $|u|<1$ (Fig.\,3,\,$b$). If we
re-expand $B(z)$ in powers of $u$,
$$
B(z)=\sum\limits_{N=0}^{\infty} \left.B_N
(-z)^N\right|_{\displaystyle z=f(u)}\qquad
\longrightarrow\qquad
B(u)=\sum\limits_{N=0}^{\infty} U_N u^N \,\,,
\eqno(7)
$$
then such series will be convergent for any  $z$. Indeed,
all singular points $P,\,Q,\,R,\ldots$ of  $B(z)$ lie on the
cut, and their images $P,\,Q,\,Q',\,R,\,R',\ldots$ in the $u$
plane appear on the boundary of the disk $|u|=1$. The
re-expanded series in (7) is convergent in the disk $|u|<1$,
but the interior of the disk is in the one-to-one correspondence
with the analyticity domain in the $z$ plane.

Such conformal mapping is unique (apart from trivial
modifications), if we want to make  analytical continuation to
the whole domain of analyticity. In fact, such strong demand is
not necessary since we need $B(z)$ only at the positive semi-axis,
in order  to produce integration in (6). If we accept that the
image of  $z=0$ is  $u=0$ and the image of  $z=\infty$ is  $u=1$,
then we can make a conformal mapping to any domain, for which the
point $u=1$ is the nearest to the origin of all boundary points
(Fig.\,3,\,$c$). The series in $u$ converges for $|u|<1$ and
particularly in the interval $0<u<1$, which is the image of the
positive semi-axis.

Such kind of conformal mapping has  advantage in the strong
coupling region. A divergency of the series in $u$ is
determined by the nearest singular point  $u=1$, which is an
image of infinity: so the large $N$ behavior of the expansion
coefficients $U_N$ is related with the strong coupling
asymptotics of $W(g)$. In order to diminish influence of other
singular points $P,\,Q,\,Q',\ldots$, it desirable to move away
these points as far, as possible. Thereby, we come
to an extremal form of such conformal mapping, when it is made on
the whole complex plane with the cut  $(1,\infty)$
(Fig.\,3,\,$d$).  Mapping of the initial region (Fig.\,3,\,$a$)
to the region of Fig.\,3,\,$d$ is given by a simple rational
transformation
$$
z=\frac{u}{a_0(1-u)} \,,
$$
for which it is easy to find the relation of  $U_N$ and $B_N$,
$$
U_0=B_0\,,\qquad U_N=\sum\limits_{K=1}^{N} \frac{B_K}{a_0^K}
   (-1)^K C_{N-1}^{K-1}\qquad (N\ge 1)\,,
\eqno(8)
$$
where $C_N^K=N!/K!(N-K)!$ are the binomial coefficients.
If $W(g)$ has a power law asymptotics
$$
W(g)=W_\infty g^\alpha \, , \qquad g \to\infty  \,,
\eqno(9)
$$
then the large order behavior of $U_N$
$$
U_N=U_\infty N^{\alpha-1}\,,\qquad N\to \infty\,,
\eqno(10)
$$
$$
U_\infty=\frac{W_\infty}{a_0^\alpha \Gamma(\alpha)
\Gamma(b+\alpha)}
\eqno(11)
$$
is determined by the parameters  $\alpha$ and $W_\infty$.
Consequently, we come to a very simple algorithm: the
coefficients  $W_N$  of the initial series  (4) define the
coefficients  $U_N$ of re-expanded series (7) according to
Eqs.\,6,\,8, while the behavior of $U_N$ at large $N$
(Eqs.\,10,\,11) is related with the strong coupling asymptotics
(9) of $W(g)$.

If information on the initial series (4) is sufficient for
establishing its strong coupling behavior (9), then
summation at arbitrary $g$ presents no problem. The
coefficients $U_N$ are calculated by Eq.\,8 for not very large
$N$, and then they are continued according to their asymptotics
(10).  Consequently, we know all coefficients of the convergent
series (7) and it can be summed with required accuracy.

One can apply the described procedure to the perturbation series
for the $\beta$-function,
$$
\beta(g)= \beta_2 g^2+\beta_3 g^3+\ldots +\beta_L g^L
+\ldots+ c_0 a_0^N \Gamma(N+b_0) g^N+ \ldots\,,
\eqno(12)
$$
having in mind that several first coefficients (till $\beta_L$)
are known from diagrammatic calculations and their large order
behavior is given by the Lipatov method. The intermediate
coefficients can be found by interpolation, the natural way for
which is as follows. It can be shown that corrections to the
Lipatov asymptotics has a form of the regular expansion in $1/N$:
$$
\beta_N=c_0 a_0^N \Gamma(N+b_0) \left\{
    1+\frac{ A_1}{N}+\frac{ A_2}{N^2}+\ldots+
    \frac{ A_K}{N^K} +\ldots  \right\}
\,.
\eqno(13)
$$
One can truncate this series and choose the retained coefficients
$A_K$ from correspondence with the first coefficients
$\beta_2,\ldots,\beta_L$; then the interpolation curve  goes
through the several known points and automatically reaches its
asymptotics.  To variate this procedure, one can re-expand the
series (13) in the inverse powers of $N-\tilde N$,
$$
\beta_N=c_0 a_0^N \Gamma(N+b_0) \left\{
    1+\frac{\tilde A_1}{N-\tilde N}+\frac{\tilde A_2}{(N-\tilde N)^2}+\ldots+
    \frac{\tilde A_K}{(N-\tilde N)^K} +\ldots  \right\}
\,,
\eqno(14)
$$
and obtain  a set of interpolations, determined by the arbitrary
parameter $\tilde N$.

In the case of four-dimensional $\phi^4$ theory, a
realization of this program \cite{109} gives the
non-alternating $\beta$-function  (Fig.\,4,\,$a$), with
the results for the exponent $\alpha$ shown in
Fig.\,4,\,$b$.
\begin{figure}
\centerline{\includegraphics[width=3.1 in]{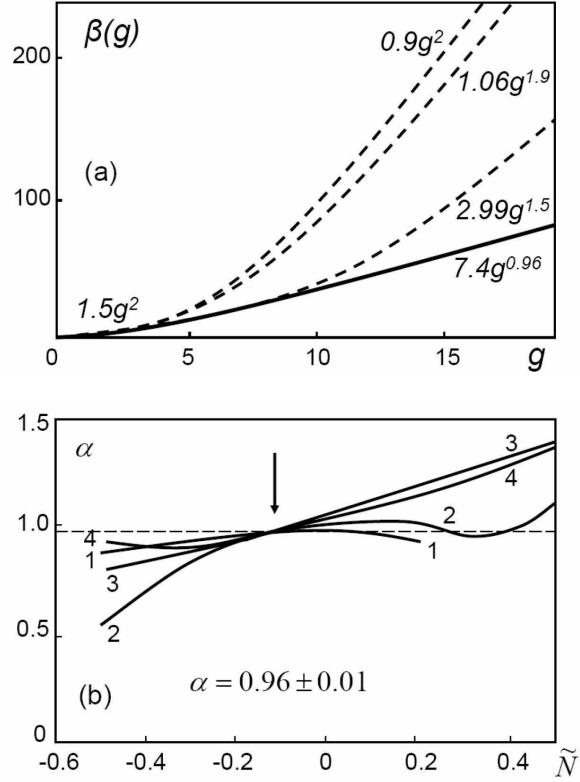}}
\caption{($a$)
General appearance of the $\beta$-function in four-dimensional
$\phi^4$ theory according to \cite{109} (solid curve), and
results obtained by other authors (upper, middle, and lower
dashed curves correspond to  \cite{203,204,205} respectively).
($b$) Different estimations of the exponent $\alpha$
according to \cite{109}. }
\label{fig4}
\end{figure}
The exponent $\alpha$ is
practically independent on $\tilde N$, and only its uncertainty
depends on this parameter. If we take the result with the minimal
uncertainty, we have a value $\alpha=0.96\pm 0.01$, surprisingly
close to unity.\,\footnote{\,Estimation of errors was made in
a framework of a certain procedure worked out in \cite{109}.
Subsequent applications have shown that such estimation is not
very reliable. }

Something close to unity is obtained also in two and three
dimensions \cite{6,7} (Fig.\,5).
\begin{figure}
\centerline{\includegraphics[width=2.5 in]{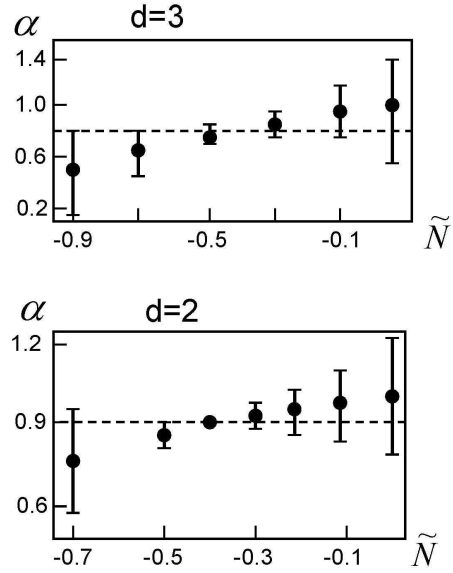}}
\caption{Estimations of the exponent $\alpha$
for $\phi^4$ theory  in two and three
dimensions \cite{6,7}. } \label{fig5}
\end{figure}
The natural
hypothesis arises, that $\beta(g)$ has the linear asymptotics
$$
\beta(g)\sim g\,\,,\qquad g\to\infty
\eqno(15)
$$
for arbitrary space dimension  $d$. If this hypothesis is
correct, then there is a natural strategy for its
justification:

(i) to test it in a simple case $d=0$;

(ii) to find out the mechanism leading to this asymptotics;

(iii) to generalize this mechanism for arbitrary $d$.

\noindent
Surprisingly, this program can be realized and Eq.\,15 is
our main result. Since summation of the series gives
non-alternating $\beta(g)$  (Fig.\,4,\,$a$), we may conclude
that the second possibility of the Bogoliubov and Shirkov
classification is realized.

\vspace{6mm}
\begin{center}
{\bf 3. "Naive" zero-dimensional limit }
\end{center}

Consider the $O(n)$-symmetric  $\phi^4$ theory with the
action
$$
S\{\phi\} =\int \,d^dx \left\{
{\textstyle\frac{1}{2}} \sum_\alpha (\nabla \phi_\alpha)^2
+ {\textstyle\frac{1}{2}} m_0^2 \sum_\alpha \phi_\alpha^{\,2} +
{\textstyle\frac{1}{8}} u
\left(\sum_\alpha\phi_\alpha^{\,2}\right)^2 \right\}\,,
$$
$$
u=g_0\Lambda^{\epsilon}\,, \qquad \epsilon=4-d
\eqno(16)
$$
in $d$--dimensional space; here $m_0$ is a bare mass, $\Lambda$
is a momentum cut-off, $g_0$ is a dimensionless bare charge.
It will be essential for us, that the $\beta$-function can be
expressed in terms of the functional
integrals\,\footnote{\,Definition of the $\beta$-function depends
on the specific renormalization scheme. We accept renormalization
conditions at zero momenta (see Sec.\,VI.\,A in \cite{8}), so the
length scale $L$ in Eq.\,3 corresponds to $m^{-1}$.}.
The general
functional integral of $\phi^4$ theory
$$
Z^{(M)}_{\alpha_1\ldots \alpha_M}(x_1,\ldots, x_M)=
%Z_M(\alpha_1,x_1,\ldots, \alpha_M,x_M) =
\int D\phi\,
\phi_{\alpha_1} (x_1) \phi_{\alpha_2} (x_2) \ldots
\phi_{\alpha_M} (x_M) \exp\left(-S\{\phi \} \right) \,
\eqno(17)
$$
contains $M$ factors of $\phi$ in the pre-exponential; this
fact is indicated by subscript $M$.

We can take a zero-dimensional limit, considering the system
restricted spatially in all directions. If its size is
sufficiently small, we can neglect the spatial dependence of
$\phi(x)$ and omit the terms with gradients in Eq.\,17;
interpreting the functional integral as a multi-dimensional
integral on a lattice, we can take the system sufficiently small,
so it contains only one lattice site. Consequently, the
functional integrals transfer to the ordinary integrals:
$$
Z^{(M)}_{\alpha_1\ldots \alpha_M} =
\int d^n\phi\,
\phi_{\alpha_1} \ldots
\phi_{\alpha_M}
\exp\left(-{\textstyle\frac{1}{2}} m_0^2 \phi^2-
{\textstyle\frac{1}{8}} u \phi^{4}   \right) \,.
\eqno(18)
$$
This is the usual understanding of zero-dimensional theory.
Such model allows to calculate any quantities with zero external
momenta. If external momenta are not zero, the model is not
complete: it does not allow to calculate the momentum
dependence. To have a closed model, we can accept that there
is no momentum dependence at
all\,\footnote{\,This point is essential in a definition of the
$Z$-factor, which should be chosen so as to give a dependence
$p^2$ with the unit coefficient in the denominator of the Green
function $G(p)$. In the described "naive" theory we accept
$Z=1$, since the momentum dependence is absent.}.
 This "naive" model is
internally consistent but does not correspond to the
true zero-dimensional limit of $\phi^4$ theory\,\footnote{\,It
will be clear below from Eq.\,49 that "naive" theory is correct
for $d=0$, while the physically interesting limit of small $d$
is singular and leads to the qualitatively different results.
}.
The latter
fact is not  essential for us, since this model is used only for
illustration and the proper consideration of the general
$d$-dimensional case will be given in the next section.

Expressing $\beta$-function in terms of functional
integrals,
we obtain it in the form of parametric representation
$$
g=1- \frac{n}{n+2} \frac{K_4 K_0}{K_2^2}
\eqno(19)
$$
$$
\beta=-\frac{2n}{n+2} \frac{K_4 K_0}{K_2^2}
\left[ 2+\frac{ \frac{K_6 K_0}{K_4 K_2}-1 }
{ 1-\frac{K_4 K_0}{K_2^2}} \right] \,.
\eqno(20)
$$
The right hand sides of these formulas contain the
integrals
$$
K_M(t) = \int_0^\infty \phi^{M+n-1} d\phi\,
\exp\left(-t \phi^2- \phi^{4}   \right) \,,
\qquad t=\left(\frac{2}{u} \right)^{1/2} \,m_0^2 \,
\eqno(21)
$$
obtained from (18) by  simple transformations.
According to (19,\,20), the quantities $g$ and $\beta$ are
functions of the single parameter $t$; excluding $t$ we
obtain the dependence $\beta(g)$.

Investigation of  (19,\,20) for real $t$ shows
that  $g$ and $\beta$ as functions of $t$ have a
behavior shown in Fig.\,6,\,$a$; a combination of these
results shows that  $\beta(g)$ behaves as in Fig.\,6,\,$b$.
\begin{figure}
\centerline{\includegraphics[width=4.5 in]{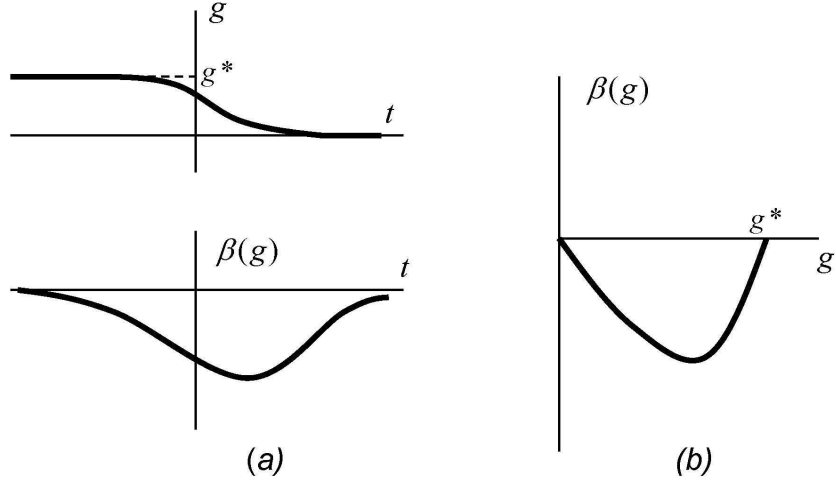}} \caption{
($a$) Dependence of  $g$ and $\beta(g)$ on the parameter $t$.
($b$) A resulting appearance of $\beta(g)$. } \label{fig6}
\end{figure}
We see that  variation of the parameter $t$ along the real axis
determines $\beta(g)$ in the finite interval $0\le g\le g^*$
where
$$
g^* = \frac{2}{n+2}\,\,.
\eqno(22)
$$
To advance into the large $g$
region, we should consider the  complex values of $t$.

It appears, that in the complex $t$ plane  we should be
interested in zeroes of the integrals $K_M(t)$. The origin
of these zeroes is very simple. There are two saddle points
in the integral $K_M(t)$, the trivial and
nontrivial,
$$
\phi_{c1}=0\,,\qquad \phi_{c2}=\sqrt{-t/2}\,,
\eqno(23)
$$
and  $K_M(t)$ can be presented as a sum of two saddle point
contributions:
$$
K_M(t)= A {\rm e}^{i\psi} + A_1 {\rm e}^{i\psi_1} \,.
 \eqno(24)
$$
If these two contributions compensate each other, then
the integral can turn to zero. Such compensation can be obtained
by adjustment of the complex parameter $t=|t|e^{i\chi}$, and in
fact there are infinite number of zeroes lying close to  lines
$\chi=\pm 3\pi/4$ and accumulating at infinity (Fig.\,7).
\begin{figure}
\centerline{\includegraphics[width=3.5 in]{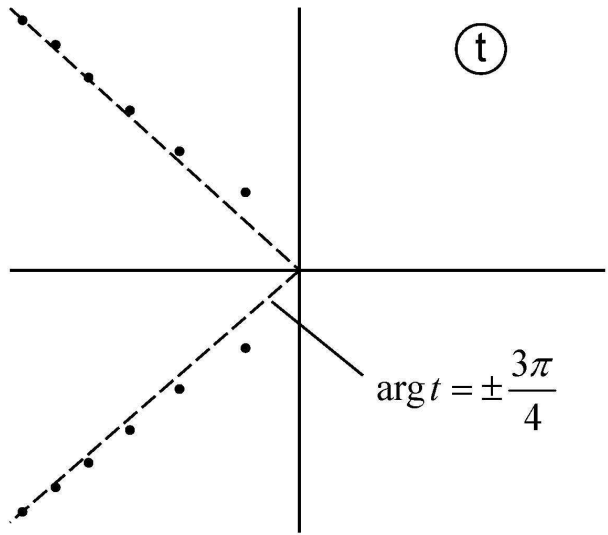}}
\caption{Zeroes of the integrals $K_M(t)$ in the complex $t$
plane. } \label{fig7}
\end{figure}
The above  saddle-point  considerations can be rigorously
justified for zeroes lying in the large $|t|$ region.
In fact, it is only essential for us that (i) zeroes of
$K_M(t)$ exist in principle, and (ii) zeroes of different
integrals  lie in different points.

Now return to the parametric representation (19,\,20). It
appears, that large values of $g$ can be achieved only
near the root of the integral  $K_2$. If $K_2$ tends to zero,
then  (19,\,20) are simplified,
$$
g\approx - \frac{n}{n+2} \frac{K_4 K_0}{K_2^2} \,,\qquad
\beta(g)\approx - \frac{4n}{n+2} \frac{K_4 K_0}{K_2^2}\,,
\eqno(25)
$$
and the parametric representation is resolved in the form
$$
\beta(g)= 4 g\,, \qquad g\to\infty \,.
\eqno(26)
$$
We see that, indeed, the asymptotic behavior of $\beta(g)$
appears to be linear.

\vspace{4mm}
\begin{center}
{\bf 4. General $d$-dimensional case} \\
\end{center}
\vspace{3mm}

The same ideas can be applied to the general $d$-dimensional
case. First of all, the actual functional integrals can turn to
zero by the same reason. Indeed,  the complex values of $t$
with large  $|t|$ correspond to complex  $g_0$ with small
$|g_0|$ (see Eq.\,21), and we come to miraculous conclusion:
large values of the renormalized charge $g$ corresponds not to
large values of the bare charge $g_0$ (as naturally to
think\,\footnote{\,It is commonly accepted that the bare charge
$g_0$ is the same quantity as the renormalized charge $g$ at the
length scale $\Lambda^{-1}$.  In fact, these two quantities
coincide only on the two-loop level \cite{9} and this relation is
valid only if $g\ll 1$ and $g_0\ll 1$ simultaneously. }),
but to its complex
values; more than that, it is sufficient to consider the region
$|g_0|\ll 1$, where the saddle-point approximation is applicable.
As a result,  the zeroes of the functional integrals can be
obtained by compensation of the saddle-point contributions of
the trivial vacuum and of the instanton configuration with the
minimal action; contributions of higher instantons are
inessential  for $|g_0|\ll 1$.

\vspace{3mm}

Now we need representation of the $\beta$-function
in terms of functional integrals. The Fourier transform of
(17) will be denoted as $K_M$ after extraction of
the $\delta$-function of momentum conservation and
a factor $I_{\alpha_1 \ldots \alpha_M}$ depending
on tensor indices:
$$
Z^{(M)}_{\alpha_1\ldots \alpha_M}(p_i)=K_M(p_i)\,
 I_{\alpha_1 \ldots \alpha_M}\,
 N\delta_{p_1+\ldots+p_M}
 \eqno(27)
$$
where $ N$ is the number of sites on the lattice,
%which is implied in definition of the functional integral,
and $I_{\alpha_1 \ldots \alpha_M}$ is a sum of terms like
$\delta_{\alpha_1\alpha_2} \delta_{\alpha_3\alpha_4} \ldots$
with all possible pairings. In general, integrals $K_M(p_i)$
are taken at zero momenta, and only the integral
$K_2$ should be known for small momentum
$$
K_2(p)=K_2-\tilde K_2 p^2+\ldots
\eqno(28)
$$
Expressing the $\beta$-function in terms of functional
integrals,
we have a  parametric representation (see \cite{114} for
details):
$$
g=-\left(\frac{K_2}{\tilde K_2} \right)^{d/2}
\frac{K_4 K_0}{K_2^2}     \,,
\eqno(29)
$$
$$
\beta=\left(\frac{K_2}{\tilde K_2} \right)^{d/2}
\left\{ -d \frac{K_4 K_0}{K_2^2}
+2 \frac{(K'_4 K_0+K_4 K'_0)K_2 -2K_4 K_0 K'_2}{K_2^2}
\frac{\tilde K_2}{K_2\tilde K'_2-K'_2\tilde K_2 }
\right\}
\eqno(30)
$$
If $g_0$ and $\Lambda$ are fixed, then the right hand sides of
these equations are functions of only $m_0$, while dependence on
the specific choice of $g_0$ and $\Lambda$ is absent due to
general theorems \cite{8}.

We see from Eq.\,29 that large values of  $g$ can be obtained
near the root of  either $K_2$, or $\tilde K_2$.
If $\tilde K_2\to 0$, equations  (29,\,30) are simplified,
so $g$ and $\beta$ are given by the same expression apart
from the factor $d$,
$$
g=-\left(\frac{K_2}{\tilde K_2} \right)^{d/2}
\frac{K_4 K_0}{K_2^2} \,, \qquad
\beta=- d \left(\frac{K_2}{\tilde K_2} \right)^{d/2}
\frac{K_4 K_0}{K_2^2}
\,,\eqno(31)
$$
and the parametric representation is resolved as
$$
\beta(g)=d g\,,\qquad g\to \infty\,.
\eqno(32)
$$
For $K_2\to 0$, the limit $g\to\infty$ can be achieved
only for $d<4$:
$$
\beta(g)=(d-4) g\,,\qquad g\to \infty\,.
\eqno(33)
$$
The results (32), (33)  correspond  to  different
branches of the analytical function  $\beta(g)$. It is easy to
understand that the physical branch is the first of them. Indeed,
it is well known from the phase transitions theory that
properties of $\phi^4$ theory change smoothly
as a function of space dimension,
and results for
$d=2,\,3$ can be obtained by analytic continuation from
$d=4-\epsilon$. According to all  available information,
the four-dimensional $\beta$-function is positive, and thus has
a positive asymptotics; by continuity, the positive asymptotics
is  expected for $d<4$.  The result (32) does obey these demands,
while
the branch (33) does not exist for $d=4$ at all.  Eq.\,32 agrees
with the approximate results discussed in Sec.\,2 and with the
exact asymptotic result $\beta(g)=2g$, obtained for
 the 2D Ising model \cite{20}
from the duality relation\,\footnote{\, Definition of the
$\beta$-function in \cite{20} differs by the sign from the
present paper.}.

%For $d=0$, Eq.\,32 does not agree with (26) by
%the reasons discussed in Sec.\,4.

\vspace{6mm}
\begin{center}
{\bf 5. Calculation of $\beta$-function in QED}
\end{center}

The same ideas can be applied to quantum electrodynamics.
Summation of the perturbation series for QED \cite{110} gives
the non-alternating $\beta$-function (Fig.\,8)
\begin{figure}
\centerline{\includegraphics[width=3.5 in]{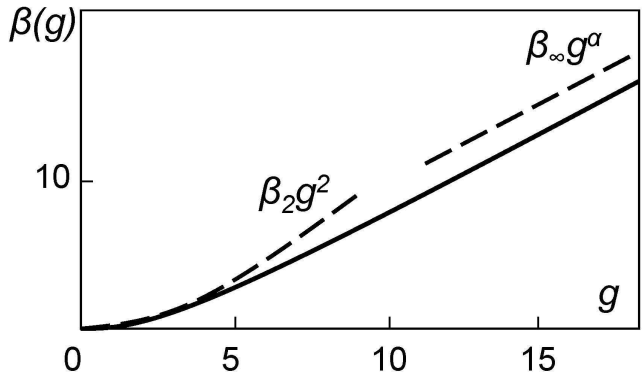}}
\caption{ General appearance of the $\beta$-function in QED
\cite{110}.} \label{fig8}
\end{figure}
with the asymptotics $\beta_\infty g^\alpha$, where (Fig.\,9)
\begin{figure}
\centerline{\includegraphics[width=3.0 in]{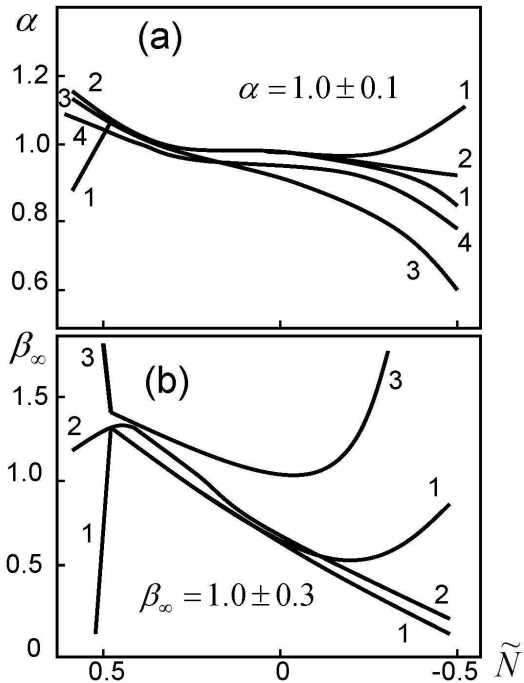}}
\caption{Different estimations of the parameters $\alpha$ and
$\beta_\infty$ for QED according to  \cite{110}. } \label{fig9}
\end{figure}
$$
\alpha=1.0\pm  0.1\,,\qquad \beta_\infty=1.0\pm 0.3 \,
\eqno(34)
$$
($g=e^2$ is the running fine structure constant).
 Within uncertainty, the obtained  $\beta$-function
satisfies inequality
 $$
 0\le \beta(g)< g \,,
 \eqno(35)
 $$
 established in \cite{11,12} from the spectral representations,
 while  asymptotics  (34) corresponds to the upper bound of
 (35).  Such coincidence is hardly incident and indicates
 that asymptotics $\beta(g)= g$ is an exact result.  We shall
see  below that it is so indeed.

The general functional integral of QED contains  $M$
photon and $2N$  fermionic fields in the
pre-exponential,
  $$
  I_{M,2N}=\int DA D\bar\psi D\psi\, A_{\mu_1}(x_1)\ldots
  A_{\mu_M}(x_M)\,
          \psi(y_1)\bar\psi(z_1)\ldots \psi(y_N)\bar\psi(z_N)
\exp\left(-S\{A,\psi,\bar\psi\}  \right) \,,
  \eqno(36)
  $$
where $S\{A,\psi,\bar\psi\}$ is the Euclidean action,
$$
S\{A,\psi,\bar\psi\} =
\int d^4x \left[ \frac{1}{4}
               (\partial_\mu A_\nu- \partial_\nu A_\mu)^2
        +\bar\psi(i\!\!\not{\! \partial} -m_0 +
    e_0 \! \not  {\!\! A})\psi \right]\,,
  \eqno(37)
$$
while $e_0$ and $m_0$ are the bare charge and mass.
%, the crossed
%symbols are convolution of the corresponding quantity with the
%Dirac matrices.
The Fourier transforms of the integrals $I_{M,N}$ with excluded
$\delta$-functions of the momentum conservation will
be referred as $K_{MN}(q_i,p_i)$ after extraction
of the usual
factors depending on tensor indices\,\footnote{\,A specific form
of these factors is inessential, since the results are
independent on the absolute normalization of $e$ and $m$.}; $q_i$
and $p_i$ are momenta of photons and electrons.

In general, these functional integrals are taken for zero
momenta, but two integrals $K_{02}(p)$ and $K_{20}(q)$ should be
estimated with the lowest order  momentum corrections: the first
is linear in $p$, and the second is quadratic in $q$,
 $$
K_{02}(p)=K_{02}+\tilde K_{02} \!\!\not{\!p}\,,\qquad
K_{20}(q)=K_{20}+\tilde K_{20} q^2\,,\qquad
\eqno(38)
$$
and in fact  the tilde denotes their momentum derivatives.

Expressing the $\beta$-function in terms of functional
integrals (see \cite{115} for details), we have a parametric
representation
$$
g= -\frac{K_{12}^2 K_{00}}{ \tilde K_{02}^2 \tilde K_{20}}\,,
\qquad
  \eqno(39)
$$
$$
\beta(g)= \frac{1}{2} \frac{K_{02} \tilde K_{02}}
{ K_{02}\tilde K'_{02} -K'_{02}\tilde K_{02} }
\frac{K_{12}^2 K_{00}}{ \tilde K_{02}^2 \tilde K_{20}}
\left\{ \frac{2\,K'_{12}}{ K_{12}} +\frac{K'_{00}}{ K_{00}}
-\frac{2\,\tilde K'_{02}}{\tilde K_{02}}
-\frac{\tilde K'_{20}}{\tilde K_{20}}
\right\}
  \eqno(40)
$$
According to  Secs.\,3,\,4, the strong coupling regime for
renormalized interaction is related with a zero of a certain
functional integral. It is clear from (39) that the limit
$g\to\infty$ can be realized by two ways:  tending to zero either
$\tilde K_{02}$ or $\tilde K_{20}$.  For $\tilde K_{02}\to 0$,
equations (39,\,40) are simplified,
$$
g= -\frac{K_{12}^2 K_{00}}{ \tilde K_{02}^2 \tilde K_{20}}\,,
\qquad
\beta(g)= -\frac{K_{12}^2 K_{00}}{ \tilde K_{02}^2 \tilde
K_{20}}\,,
\eqno(41)
$$
and the parametric representation is resolved in the form
$$
\beta(g)= g\,,\qquad g\to \infty\,.
\eqno(42)
$$
For $\tilde K_{20}\to 0$, one has
$$
\beta(g)\propto g^2\,,\qquad g\to \infty\,.
\eqno(43)
$$
Consequently, there are two possibilities for the
asymptotics of  $\beta(g)$, either (42) or (43). The
second possibility is in conflict with inequality (35),
while the first one is in excellent agreement with
results (34) obtained by summation of the perturbation series.
In our opinion, it is a sufficient reason to consider
(42) as an exact result for asymptotical behavior of the
 $\beta$-function. It means that the fine structure
 constant in pure QED behaves as  $g\propto L^{-2}$ at small
 distances.

\vspace{6mm}
\begin{center}
{\bf 6. A scheme without complex parameters}
\end{center}

Our use of the complex bare parameters may look suspicious,
since it  corresponds to the non-Hermitian bare Hamiltonian; at
first glance, it violates unitarity, since the $S$-matrix is
expressed through the Dyson $T$-exponential of the bare action.

In fact, a problem is solved by Bogoliubov's construction of
the axiomatical $S$-matrix \cite{104}:
according to it, the general form of the $S$-matrix is
given by the $T$-exponential of $iA$, where $A$ is a sum of (i)
the bare action, and (ii) a sequence of arbitrary "integration
constants" which are determined by quasi-local operators.  In the
regularized theory we can set the "integration constants" to be
zero, and the $S$-matrix is determined by the bare action.
However, in the course of renormalization these constants are
taken non-zero, in order to remove divergences.  These non-zero
"integration constants" can be absorbed by the action due to the
change of its parameters. As a result, for the true continual
theory the $S$-matrix  is determined by the renormalized
Lagrangian, which is Hermitian for real $g$.

Nevertheless, scientific community has a bias against complex
bare parameters: it is related with the old discussion between
Lee and Pauli, moderated by  Heisenberg, on the exactly
solvable model suggested by Lee \cite{xx1}. After paper
\cite{xx4}, the Lee model was considered as unsatisfactory due to
existence of "ghost states", and this point of view was included
in many textbooks. Quite recently \cite{xx5} it was found that
this point of view is incorrect and the Lee model is completely
acceptable. A key idea of \cite{xx5} is that the
complex-valued Hamiltonian can be made Herminian by modification
of the inner product for the corresponding Hilbert space.

Below we can calm sceptic spirits and suggest a scheme without
complex parameters.   In this case we accept explicitly the
lattice regularization and take the action in the form
$$
S\{\phi\} ={\textstyle\frac{1}{2}} a^d
\sum_{\bf x,x'} J_{\bf x-x'} \phi_{\bf x} \phi_{\bf x'}
+ {\textstyle\frac{1}{2}} m_0^2 a^d
                 \sum_{\bf x}  \phi_{\bf x}^2
+{\textstyle\frac{1}{4}} g_0 a^{2d-4}
         \sum_{\bf x}  \phi_{\bf x}^4 \,,
\eqno(44)
$$
where we accept $\Lambda=a^{-1}$ ($a$ is a lattice spacing)
and restrict ourselves by the case $n=1$. Making a change of
variables
$$
\phi \longrightarrow \phi
\left(g_0 a^{2d-4}/4 \right)^{-1/4}
\eqno(45)
$$
and setting  $t=(1/g_0)^{1/2}$ as before, we can write the
functional integral (17) in the form
$$
Z^{(M)}\{{\bf x}_i\}=(2t)^{\frac{{\cal
N}+M}{2}} \int \left(\prod_{\bf x}\,d\phi_{\bf x} \right)
      \phi_{{\bf x}_1} \ldots \phi_{{\bf x}_M}
\exp\left\{-t\sum_{\bf x,x'} J_{\bf x-x'}
\phi_{\bf x}\phi_{\bf x'}
-t m_0^2 \sum_{\bf x}  \phi_{\bf x}^2
-\sum_{\bf x}  \phi_{\bf x}^4
\right\}.
\eqno(46)
$$
We accept $a=1$ measuring $J_{\bf x-x'}$ and $m_0^2$ in units of
$\Lambda^2$. We consider  $t$ as a running parameter of the
parametric representation and investigate a singularity at
$t\to 0$, which has a simple origin. For $g_0\gg 1$, Eq.\,46
allows expansion over the gradient term
$t J_{\bf x-x'} \phi_{\bf x}\phi_{\bf x'}$. In zero order
in $t$ the integral  $Z^{(2)}$ has a $\delta$-functional
form in the coordinate representation,
$Z^{(2)}({\bf x},{\bf x'})\sim \delta_{{\bf x}{\bf x'}}$,
and its Fourier transform has no momentum dependence;
the latter appears only in the first order in $t$. As a result,
the integral $\tilde K_2$ in expansion (28) is small in
comparison with $K_2$, i.e. $K_2/\tilde K_2\sim 1/t$, which
leads to a singularity at $t=0$ in (29,\,30). This singularity
is more complicated than other singularities in the $t$ plane
(Fig.7) and needs the accurate investigation.

Expansion of the functional integral (46)  in powers of $t$
expresses it in terms of ordinary integrals
$$
I_{2k} =\int_{-\infty}^{\infty} d\phi \,
  \phi^{2k}
\exp\left\{-t m_0^2 \phi^2 - \phi^4  \right\} \,
\eqno(47)
$$
and parametric representation (29,\,30) reduces to the form
$$
g=\left( \frac{n}{2t}\,\frac{I_0}{I_2} \right)^{d/2}
\left(1- \frac{n}{n\!+\!2}\frac{I_4 I_0}{I_2^2} \right)
  \,,
$$
$$
\frac{\beta(g)}{g}=
 d +2 \,\frac{\displaystyle \frac{I_6 I_2}{I_0^2}
    -\frac{ 2 I_4^2}{I_0^2} +\frac{ I_2^2 I_4}{I_0^3} }
{ \displaystyle\left(\frac{I_4}{I_0}- \frac{n\!+\!2}{n}\frac{I_2^2}{I_0^2}
\right)
  \left( \frac{I_2^2}{I_0^2} - \frac{I_4}{I_0}\right)}
    \,.
\eqno(48)
  $$
This result can be written in a simple form, if the
functions  $g_{zero}(t)$ and $\beta_{zero}(t)$ are introduced,
which correspond to the zero-dimensional case and have
appearance shown in Fig.6,$a$:
$$
g=\left( \frac{n}{2t}\,\frac{I_0}{I_2}
\right)^{d/2} g_{zero}(tm_0^2)
  \,,
$$
$$
\beta(g)=\left( \frac{n}{2t}\,\frac{I_0}{I_2}
\right)^{d/2} \left[d g_{zero}(tm_0^2)+\beta_{zero}(tm_0^2)
\right] \,,
\eqno(49)
$$
Resolving the parametric representation in the limit
$t\to 0$, we come to the asymptotics
$$
\beta(g)=\left[d+\frac{\beta_{zero}(0)}{g_{zero}(0)}
\right]\, g \,, \qquad g\to\infty
$$
reducing to the result $\beta(g)=2.29 \,g$ of the paper
\cite{415} after substitution of numerical values.
However, this result is not final. Instead of the limit
$t\to 0$ for fixed $m_0$ one can consider a limiting
transition under condition  $t m_0^2\to const$ with different
values of $const$; then the general structure of theory remains
unchanged, but asymptotical behavior of $\beta(g)$ will be
different. The question arises on the correct character of the
limiting transition, corresponding to the strong coupling regime.

The gradient expansions of the considered type was exploited
in a number of works \cite{415}-\cite{416}, and ambiguity of
a strong coupling limit was finally realized by their authors.
Nevertheless, the correct character of the limiting
transition was not established till the paper \cite{116}.
In the framework of the parametric representation (29,\,30)
the indicated problem accepts the different form. Deficiency
of the result (49) consists in the presence of two independent
parameters  $t$ and $tm_0^2$. If one of them is
excluded in favor of $g$, then the $\beta$-function depends
not only on $g$ but also on $m_0^2$, while the latter
dependence should be absent according to general theorems
\cite{8}. The question arises on resolving of this
contradiction.
  Of course, there is no real contradiction, because the
general theorems suggest that the continual limit
$\Lambda\to\infty$ is already taken.  Physically it means a
fulfilment of the condition
$$
m^2\ll \Lambda^2 \,,
\eqno(50)
$$
which is equivalent to the condition $\xi\gg a$ for the
correlation length $\xi$; it means that a characteristic scale
of the field variation contains many lattice sites, so
further diminishing of $a$ is of no significance.
\begin{figure}
\centerline{\includegraphics[width=4.0 in]{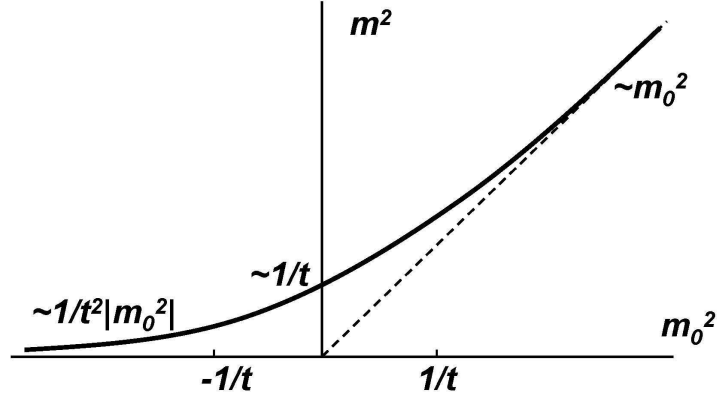}}
\caption{The renormalized mass as a function of the bare
mass in the strong coupling region. } \label{fig10}
\end{figure}
As a  result  of the gradient expansions, the renormalized
mass is represented in the form (Fig.10)
$$
m^2=\frac{K_2}{\tilde K_2}= \frac{n}{2t}\,\frac{I_0}{I_2}
= \left\{ \begin{array}{cc}
m_0^2\,, & t m_0^2\gg 1\\
{ }\\
\sim 1/t\,, &   |t m_0^2| \alt 1
\\ {   }\\
\sim 1/t^2 |m_0^2|  \,, & -t m_0^2\gg 1 \,,
\end{array} \right.
\eqno(51)
$$
and the condition $m^2\ll 1$ (corresponding to (50) in dimensional
units) is satisfied in the case
$$
t m_0^2 = -\kappa\,, \qquad \kappa\gg 1\,.
\eqno(52)
$$
After replacement $\phi_{{\bf x}}^2 \to
\kappa\phi_{{\bf x}}^2 /2$ the exponential in (46) accepts  a
form
$$
\exp\left\{-{\textstyle\frac{1}{2}} t\kappa\,\sum_{\bf x,x'} J_{\bf x-x'}
\phi_{\bf x}\phi_{\bf x'}
\right\}
 \prod_{\bf x}
\exp\left\{ {\textstyle\frac{1}{4}} \kappa^2
\left(2\phi_{\bf x}^2 -  \phi_{\bf x}^4 \right) \right\}
\eqno(53)
$$
where the last factor is localized near $\phi^2_{\bf x}=1$
and can be replaced by $A\delta(\phi^2_{\bf x}-1)$. The
constant $A$ is inessential for a ratio of two
integrals and one can set  $A=1$. As a result, Eq.\,46
accepts a form
$$
Z_M\{{\bf x}_i\}=(t\kappa)^{\frac{{\cal N}+M}{2}}
    \int \left(\prod_{\bf x}\,d\phi_{\bf x} \right)
      \phi_{{\bf x}_1} \ldots \phi_{{\bf x}_M}
\exp\left\{-{\textstyle\frac{1}{2}} t\kappa\,\sum_{\bf x,x'}
J_{\bf x-x'} \phi_{\bf x}\phi_{\bf x'} \right\} \prod_{\bf
 x} \delta(\phi^2_{\bf x}-1)
\eqno(54)
$$
and a functional integral transforms into the Ising sum over
values  $\phi_{\bf x}=\pm 1$. In the $n$-component case one
obtains a $\sigma$-model \cite{301} instead of the Ising model.

Now all functional integrals depend on the single
variable $t\kappa$ and the right-hand sides of (29,\,30)
are functions of only this argument; it  defines the
$\beta$-function depending only on $g$. In the physical
motivation of the limiting transition we implied the conditions
providing (50)
$$
t\ll 1\,,\qquad \kappa\gg 1\,, \qquad t\kappa\gg 1\,,
\eqno(55)
$$
but in fact only first two inequalities  were used in
derivation of (54). Therefore, transformation to the Ising
model is valid under conditions
$$
t\ll 1\,,\qquad \kappa\gg 1\,, \qquad t\kappa\quad \mbox{is
arbitrary}\,.
\eqno(56)
$$
In particular, it is valid  in the region
$t\kappa\ll 1$, where  gradient expansions are possible and
large values of the renormalized charge  $g$ are reached.
Using (51,\,52), we come to conclusion that strong coupling regime
of $\phi^4$ theory corresponds to the limit
$$
t\to 0\,,\qquad t m_0^2 \to -\infty\,,
\qquad t m^2 \to 0\,, \qquad m^2\to\infty
\,,
$$
so neither $m_0=const$ nor $m=const$ is the correct condition
for a limiting transition. It means that dependencies of Fig.2 are
not actual from the very beginning.

Returning to (49) and setting  $-t m_0^2=\kappa\gg 1$,
we have
$$
g=\left( \frac{n}{t\kappa }\right)^{d/2} g^* \,, \qquad
\beta(g)=\left( \frac{n}{t \kappa}\,\right)^{d/2} dg^* \,,
\eqno(57)
$$
and so the asymptotics of the $\beta$-function is obtained
$$
\beta(g)=d g\, \qquad (g\to  \infty)\,,
 \eqno(58)
 $$
coinciding with (32). We see that singularity at $t=0$ leads
to the same result as singularities at the complex values of $t$.

Representation (54) for functional integrals can be used for
calculation of the observable quantities. The latter are
obtained in the form
$$
A_{obs}= \Lambda^{d_A} f_A(t\kappa)\,,
\eqno(59)
$$
where $d_A$ is a physical dimension of the quantity $A_{obs}$.
Using analogous expressions for $g$ and  $m$,
$$
g=f_g(t\kappa)\,,\qquad
m^2 = \Lambda^2 f_m(t\kappa) \,,
\eqno(60)
$$
we can rewrite (59) in the form
$$
A_{obs}= m^{d_A} F(g) \,,
\eqno(61)
$$
which does not contained the bare parameters $g_0$, $m_0$,
$\Lambda$; so (61) gives a "theorem of renormalizability"
for a strong coupling region. It should be stressed that we do
not take the continual limit in the bare theory and retain the
lattice as a convenient instrument for representation
of functional integrals, and only the lattice spacing $a$
is excluded from the physical results.

\begin{center}
{\bf 7. Application of high temperature expansions}
\end{center}

Rewriting (56) in dimensional quantities and changing
$t\kappa\to \kappa$, we see that reducing of $\phi^4$ theory to
the Ising model is possible under conditions
$$
 g_0\gg 1\,,\qquad -g_0^{-1/2} m_0^2/\Lambda^2 \gg 1\,,
\qquad -g_0^{-1} m_0^2/\Lambda^2 =\kappa
 \,,
\eqno(62)
$$
where $\kappa$ is an arbitrary parameter having a sense of the
inverse temperature in the Ising model. Correspondingly, Eqs.\,48
have a structure
$$
g=F(\kappa)\,,\qquad
\beta(g) = F_1(\kappa) \,,
\eqno(63)
$$
and define the $\beta$-function in the parametric form. At first
glance, the condition $g_0\gg 1$ corresponds to the strong
coupling regime and parametric representation (63) is limited
only by this regime. However, there is another view on this
situation. Let strengthen conditions (62) by taking the
limit
$$
g_0\to\infty\,,\qquad -g_0^{-1/2} m_0^2/\Lambda^2 \to \infty\,,
\qquad -g_0^{-1} m_0^2/\Lambda^2 =\kappa
\eqno(64)
$$
In this case, transition from (46) to (54)
is valid without any approximations and conserves a
strict equivalence with the initial $\phi^4$ theory under a
certain choice of its bare parameters. The last property
conserves a form of the Lagrangian
under renormalizations. The taken limit
$g_0\to\infty$ does not mean the same limit for the
renormalized charge $g$; in fact, according to gradient
expansions, $g$ varies from infinity to the order of unity
when $\kappa$ changes from zero to a finite value.  Since
parametric representation (63) is exact and specifies the $\beta$
function in the interval $1\alt  g <\infty$, it can be
analytically continued and treated as a definition of $\beta(g)$
at arbitrary values of $g$. However, a certain doubt arises:
does this definition provide the correct results in the
weak-coupling region?

An answer to this question can be obtained using high-temperature
series \cite{605}. Such series are traditionally constructed for
quantities
(superscript $c$ marks the connected diagrams)
$$
\raisebox{2pt}{$\chi$}_2=\sum_{\bf x} \langle \phi_{{\bf
x}}\phi_{{\bf 0}} \rangle^c\,,\qquad
\raisebox{2pt}{$\mu$}_2=\sum_{\bf x} {\bf x}^2\langle
\phi_{{\bf x}} \phi_{{\bf 0}} \rangle^c\,,\qquad
\raisebox{2pt}{$\chi$}_4=\sum_{\bf x,y,z} \langle
\phi_{{\bf x}} \phi_{{\bf y}}\phi_{{\bf z}}\phi_{{\bf
0}} \rangle^c\,,\qquad
\eqno(65)
$$
which coincides up to factors with the ratios $K_2/K_0$,
$\tilde K_2/K_0$, and $K_4/K_0$ of the functional integrals
specified above; more precisely,
$$
\frac{K_2}{\tilde K_2}\,=\,2d\, \frac{\raisebox{2pt}{$\chi$}_2}
{\raisebox{1pt}{$\mu$}_2}\, \equiv
\frac{1}{\kappa} f_0(\kappa) \,,
$$
$$
\frac{K_2}{K_0}\,=\,2\kappa\,\raisebox{2pt}{$\chi$}_2
\equiv \kappa f_2(\kappa) \,,
\eqno(66)
$$
$$
\frac{K_4 K_0}{K_2^2}\,=\,\frac{1}{3}\,
\frac{\raisebox{2pt}{$\chi$}_4}{\raisebox{2pt}{$\chi$}_2^2}
\,\equiv - f_4(\kappa) \,,
$$
where the introduced functions $f_i(\kappa)$ will be used below.
It was taken into account that there is no zeroth term in the
expansion of ${\raisebox{1pt}{$\mu$}_2}$ in $\kappa$
(see Eq.\,68 below), so that all functions $f_0(\kappa)$,
$f_2(\kappa)$, and $f_4(\kappa)$ are regular and their
expansions begin with the zeroth term. The substitution of
(66) into (29,\,30) gives
$$
g=\left(\frac{f_0(\kappa)}{\kappa} \right)^{d/2}
f_4(\kappa) \,,  \qquad
\frac{\beta(g)}{g}= d - 2\kappa \,\frac{[\ln f_4(\kappa)]'  }
{ 1-\kappa \,[\ln f_0(\kappa)]' }    \,.
\eqno(67)
$$
The initial parametric representation (29,\,30) for the
$\beta$-function is "dead", since evaluation of functional
integrals looks hopeless. It becomes "alive" in the form
(67), since functions $f_i(\kappa)$ can be calculated using
high-temperature expansions\,\footnote{\,The function
$f_2(\kappa)$ does not enter to (67), but it is actual for
calculation of anomalous dimensions \cite{117} }.

For a simple hypercubic lattice with the
interaction between the nearest neighbors, the first terms of
the expansion for functions (65)  have  the form (in the case
$d = 4$,  $n = 1$)  \cite{617}
$$
\raisebox{2pt}{$\chi$}_2= 1 +16 \kappa + 224 \kappa^2 + \ldots
$$
$$
\raisebox{1pt}{$\mu$}_2= 16 \kappa + 512 \kappa^2 +
33920/3 \kappa^3 +\ldots
\eqno(68)
$$
$$
\raisebox{2pt}{$\chi$}_4= -2 - 128 \kappa - 4672 \kappa^2 -
\ldots
$$
Taking limit $\kappa\to 0$, it is easy to obtain the strong
coupling behavior for the $\beta$-function and anomalous
dimensions \cite{117}, and then develop their expansion in powers
of $g^{-2/d}$.  Below, 14 terms of expansion (68) are used,
which are given for $n = 1$ in tables 5, 8, and 11 of the
paper \cite{617}.

The use of Pade-approximants allows to obtain the
$\beta$-function and anomalous dimensions at arbitrary $g$.
The general strategy consists in the following. The Ising model
has a phase transition in a certain point $\kappa_c=1/T_c$, and
a typical physical quantity $F(\kappa)$ has the critical
behavior of the form
$$
F\propto (T-T_c)^{-\lambda} \propto
(\kappa_c-\kappa)^{-\lambda} \,.
\eqno(69)
$$
If $F(\kappa)$ is expanded in $\kappa$,  then a convergence
radius of the expansion is limited by the quantity
$\kappa_c$; in actual cases, $\kappa_c$ is the nearest
singularity to the coordinate origin. If the logarithmic
derivative of $F$ is taken,
$$
(\ln F)' =\frac{F'}{F} = \frac{{-\lambda}}{\kappa-\kappa_c}
+ \mbox{less singular terms}  \,,
\eqno(70)
$$
then the main singularity  for it
is a simple pole with a residue ${-\lambda}$ and can be
investigated using the Pade approximation \cite{618}. The
Pade-approximant $[M/N]$ is defined as the ratio of two
polynomials of  degrees $M$ and $N$,
$$
(\ln F)' = \,\frac{P_M(\kappa)}{Q_N(\kappa)}\,= \,
\frac{p_0+p_1 \kappa+\ldots +p_M \kappa^M}
{1+q_1 \kappa+\ldots +q_N \kappa^N} \,\,,
\eqno(71)
$$
whose coefficients are chosen to reproduce  the first $M +N+ 1$
coefficients in the expansion of $(\ln F)'$ over $\kappa$.  It is
known that Pade-approximants successfully predict the nearest
singularities of the corresponding function if these singularities
are the simple poles \cite{605,618}\,\footnote{\,Usually, one uses
diagonal ($M = N$) or quasi-diagonal ($M\approx N$) approximants,
whose convergence to the corresponding function is proved under
the most general assumptions \cite{618}.}. If $\lambda$ and
$\kappa_c$ are predicted reliably, then the whole function
$F(\kappa)$ can be found in the interval $0\le \kappa\le \kappa_c$
with a good precision. If such results for $f_i(\kappa)$ are
substituted into the right-hand sides of (67), then the
$\beta$-function can be determined in the interval $ g^*\le  g
<\infty$, where $g^*$ is the fixed point of the renormalization
group. In the four-dimensional case, one has $g^* =0$ and
$\beta(g)$ is completely determined by the described procedure.

The use of this strategy in the four-dimensional case is
complicated by the existence of logarithmic corrections
to scaling \cite{619,8}:
$$
\raisebox{2pt}{$\chi$}_2 \sim \tau^{-1} |\ln \tau|^p \,,
\qquad \xi^2 \sim \frac{\raisebox{1pt}{$\mu$}_2}
{\raisebox{2pt}{$\chi$}_2}
 \sim \tau^{-1} |\ln \tau|^p\,,
\qquad \raisebox{2pt}{$\chi$}_4 \sim \tau^{-4} |\ln
\tau|^{4p-1}\,,
\eqno(72)
$$
where $\tau\sim (\kappa_c-\kappa)$ is a distance to the
transition and  $p=(n+2)/(n+8)$. Substitution
to (66,\,67) gives for  behavior of charge $g$
$$
g=\frac{c_0}{|\ln \tau|} \,, \qquad c_0=2/\beta_2 \qquad
(\tau\to 0)  \,,
\eqno(73)
$$
where the coefficient of the logarithmic factor is universal.
If equations (72) and (73) are fulfilled, then the parametric
representation (67) automatically reproduces the one-loop result
of weak-coupling expansion for the $\beta$-function.

The objective test of Eqs.\,72 for lattice models were performed
in many works \cite{606}--\cite{616}. In particular, it was
convincingly shown in \cite{606,607} that high-temperature series
for the Ising model allow reliable prediction of the exponent
$p$.  Eq.\,73 was confirmed with a satisfactory accuracy in
\cite{607,609}.  Already these results provide the positive
answer to the question formulated above:  parametric
representation (63) gives correct results for the
$\beta$-function in the weak-coupling region.

The modified treatment procedure suggested in \cite{117} allows to
improve estimation of the constant $c_0$, which was not very
satisfactory in preceding papers. The relation
$\tau=A_0(1-\kappa/\kappa_c)$ contains the non-universal
coefficient $A_0$ which is essential in logarithmic factors. If
such factors are extracted from functions (65) in accordance with
(72) and the Pade analysis is applied to remaining functions, then
strong coupling estimations of the constant $c_0$ (Fig.11,$a$) are
close to the
\begin{figure}
\centerline{\includegraphics[width=6.0 in]{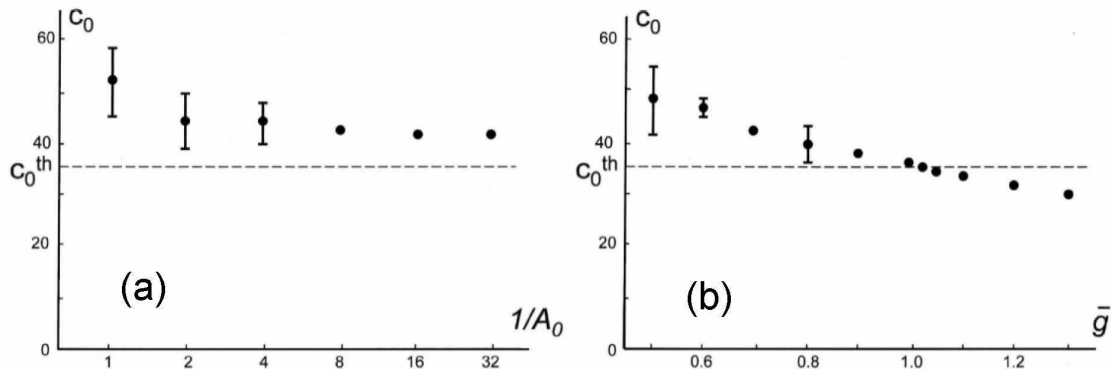}}
\caption{The constant $c_0$ in Eq.\,73  versus   $A_0$
 in the leading logarithmic approximation ($a$), and  versus
$\bar g$ in the next-to-leading logarithmic approximation
($b$). }
 \label{fig11}
\end{figure}
theoretical value
$c_0^{th}=35.09$  but systematically exceed it.  If the
logarithmic factors are extracted in accordance with the
next-to-leading logarithmic approximation, then estimations of
$c_0$ improve (Fig.11,$b$) and  become close to the theoretical
value in the same range $0.85\div 1.06$  of the non-universal
parameter $\bar g$ (analogous to $A_0$) where the power law
singularities of remaining factors are close to theoretical
expectations.  The exact $c_0$ value is realized  at $\bar
g\approx 1.02$, and this value of $\bar g$ can be used in the
further analysis.  Extracting from the quantities
$\raisebox{2pt}{$\chi$}_2$, $\raisebox{2pt}{$\mu$}_2$,
$\raisebox{2pt}{$\chi$}_4$ all logarithmic and power-law
singularities and applying the Pade-approximation to remaining
regular functions, one can calculate the right hand sides of
(67)  for the whole interval $0<\kappa<\kappa_c$ and obtain the
$\beta$-function at arbitrary $g$ (see a solid line in Fig.12).
To give an impression on accuracy of the calculation, we show by
the dotted line the analogous dependence obtained under the
assumption of the constancy of indicated regular functions, when
any information on them is dropped from results. In fact, these
regular functions are known with the accuracy of several
percents.
\begin{figure}
\centerline{\includegraphics[width=5.0 in]{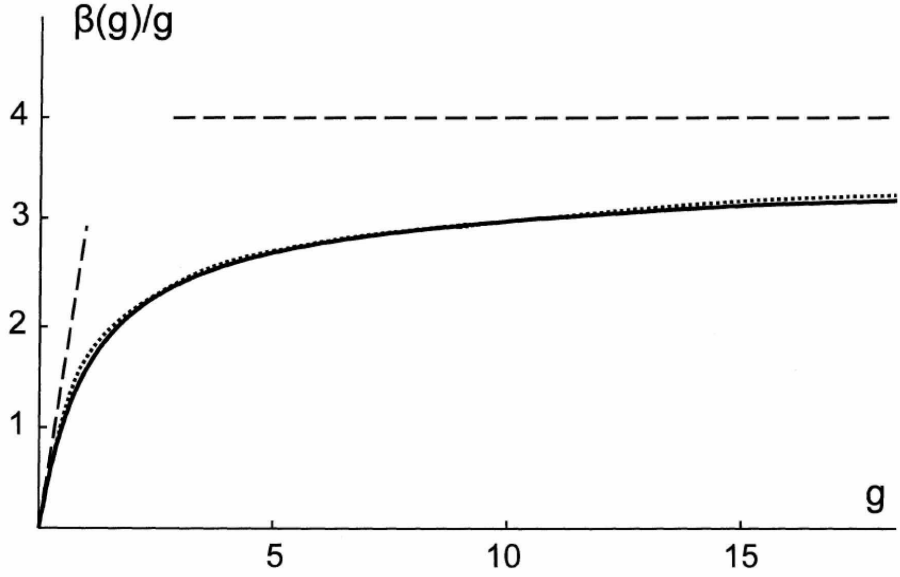}}
\caption{A solid line is a calculated function $\beta(g)/g$.
The dashed lines are the strong- and weak-coupling asymptotic
behaviors. A dotted line shows the same results obtained under
the assumption of the constancy of regular functions.}
\label{fig12}
\end{figure}

Analogously, one can obtain a behavior of the renormalized charge
$g$ and the renormalized mass $m$ as functions of $\kappa$
(Fig.13), which is essential for characterization of the lattice
$\phi^4$ theory.
\begin{figure}
\centerline{\includegraphics[width=5.1 in]{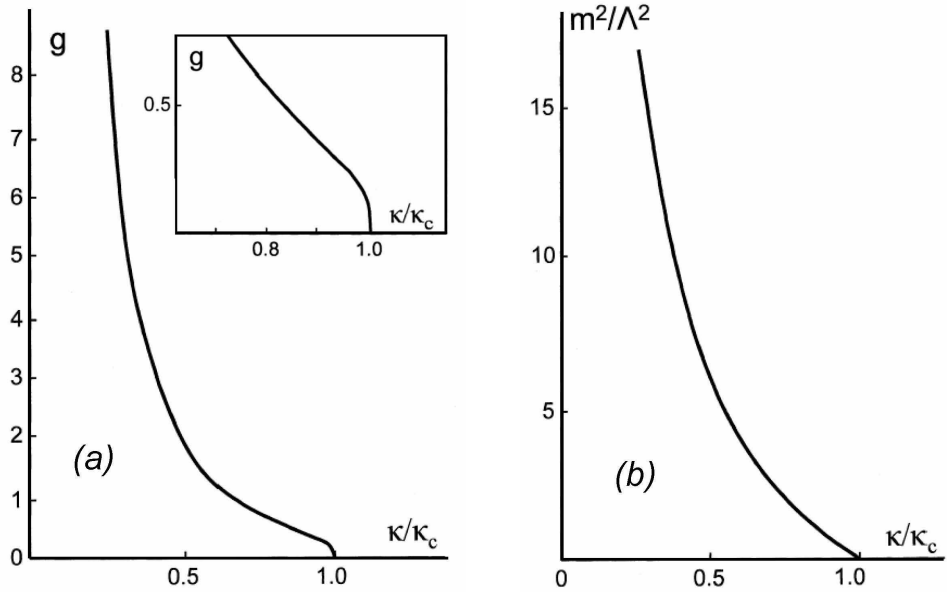}}
\caption{Renormalized  charge $g$ ($a$) and  mass $m$ ($b$) versus
$\kappa/\kappa_c$. }
\label{fig13}
\end{figure}

\begin{center}
{\bf 8. Is $\phi^4$ theory trivial? }
\end{center}

In the preceding sections we have established, that the Gell-Mann
-- Low function $\beta(g)$ in four-dimensional $\phi^4$ theory is
non-alternating and has asymptotic behavior  $\beta(g)=4 g$ at
$g\to\infty$. According to the Bogoliubov and Shirkov
classification (Sec.\,1), it means a possibility to construct the
continuous theory with finite interaction at large distances.
This conclusion is in visible contradiction with lattice results
indicating triviality of $\phi^4$ theory.

As we stressed in Sec.\,1, one should differ two definitions
of triviality. Wilson triviality means
that integration of equation (3) in the direction of large
distances $L$ gives the effective charge $g$ tending to zero;
this definition implies the massless theory, since in the
opposite case the distance scale is saturated by the inverse
mass. In definition of true triviality   one considers the
massive theory and suggests  finite interaction $g_{\infty}$
for $L\agt m^{-1}$; a theory is trivial, if integration of (3) in
direction of small $L$ gives a divergency at finite $L_0$
and does not allow to reach the $L\to 0$ limit.  Such situation
is internally inconsistent and means incorrectness of the initial
suggestion on finite interaction at large distances; in fact,
$L_0\to 0$ if $g_{\infty}\to 0$. Wilson triviality means
that $\beta$-function is non-negative and has a zero only for
$g=0$. True triviality needs in addition its sufficiently
quick growth at infinity, $\beta(g)\sim g^\alpha$ with
$\alpha>1$. According to preceding sections, $\phi^4$ theory
and QED are trivial in the Wilson sense, but do not possess
true triviality.

\vspace{2mm}

Two definitions of triviality were hopelessly mixed in
literature \cite{113}, and there are two main reasons for that.
Firstly, it is rather difficult to test true triviality in
the lattice approach\,\footnote{\,A definition of
true triviality in the lattice approach was
given in mathematical papers  \cite{106,107}. When the lattice
spacing  $a$ tends to zero, the bare parameters $g_0$
and $m_0$ should be considered as functions of $a$.  A theory is
non-trivial, if there exists some choice of functions $g_0(a)$ and
$m_0(a)$, providing finite interaction at large $L$; if
such functions do not exist, then a theory is trivial. Of course,
it is rather difficult to test "existence" or "non-existence" in
numerical simulations.}. Secondly, one can advance arguments
that give an illusion of equivalence of two definitions.
As illustration  to the latter point, consider the following
reasoning.  The only alternative to perturbative approach is to
express all quantities  in terms
of the functional integrals. The latter depend on the
bare charge $g_0$, bare mass $m_0$ and the ultraviolet cut-off
$\Lambda$. Taking into account their dimensional
character, one has the following relations for the
renormalized charge $g$, renormalized mass $m$ and
observable quantities $A_i$
$$
g=F_g\left( g_0, m_0/\Lambda\right)\,,\qquad
$$
$$
m=\Lambda F_m\left( g_0, m_0/\Lambda\right)\,,\qquad
\eqno(74)
$$
$$
A_i=\Lambda^{d_i} F_i\left( g_0, m_0/\Lambda\right)\,,\qquad
$$
where $d_i$ is a physical dimensionality of $A_i$.
Excluding  $g_0$ and $m_0/\Lambda$ in favor of $g$ and
$m/\Lambda$, one has
$$
A_i=m^{d_i} \tilde F_i\left( g, m/\Lambda\right)\,.\qquad
\eqno(75)
$$
To eliminate the dependence on $\Lambda$ we should take the
limit $m/\Lambda\to 0$. In the lattice approach, this limit
corresponds to $\xi/a\to\infty$, i.e. to the phase
transition point. The latter is determined by a zero of
$\beta$-function, which gives $g=0$  in four-dimensional
$\phi^4$ theory.

In this argumentation,  Wilson triviality is considered as
given, while  true triviality is "derived" from it. Of course,
such "proof" is incorrect, because two definitions
are surely not equivalent.  This shortcoming originates from our
assumption that a  generic situation takes place in
Eq.\,75, and a limit $m/\Lambda\to 0$ is necessary. However, the
special case is possible, when the $m/\Lambda$ dependence
is absent in (75), and a limiting transition is irrelevant.
This special case fills the "gap" between two definitions and
makes them not equivalent.

According to Sec.\,6, such special case is actually realized
in $\phi^4$ theory. Let return to Eqs.\,74 and
impose the condition $m \ll \Lambda$, corresponding to the
continuum limit of  renormalized theory. If this condition is
imposed in the region $g_0\gg 1$, then  $\phi^4$ theory reduces
to the Ising model, containing the single parameter  $\kappa$,
which plays the role of inverse temperature;  relations
(74) accept the form
$$
g=F_g\left( \kappa\right)\,,\qquad
$$
$$
m=\Lambda F_m\left( \kappa\right) \,,\qquad
\eqno(76)
$$
$$
A_i=\Lambda^{d_i} F_i\left(\kappa\right)\,. \qquad
$$
So far there is nothing unusual: the condition
$m/\Lambda\to 0$ gives a relation between $g_0$ and
$m_0/\Lambda$, so all functions in Eq.\,74 depend on
the single  parameter,  which we denoted as  $\kappa$.
The non-trivial point consists in the fact that condition
$m/\Lambda\ll 1$ is sufficient for transformation to the Ising
model, but not necessary for it.  This transformation is
possible under the weaker conditions, which are compatible with
an arbitrary value of $m/\Lambda$ (Sec.\,6). Excluding $\kappa$
from (76), one obtains the equations
$$
A_i=m^{d_i}  F_i\left( g \right)\,,\qquad
\eqno(77)
$$
which are
analogous to (75), but do not contain the parameter $m/\Lambda$.
As a result, the program of renormalization is completely
fulfilled, and no additional limiting transitions are necessary.
It means that (a) we can retain the lattice in the bare theory
(as a convenient tool for representation of functional
integrals), and (b) relation between  $m$ and $\Lambda$ (or $\xi$
and $a$) can be arbitrary, so a finite value of $g$ becomes
possible.

Usually, the lattice theory  contains more parameters than
the initial field theory. For example, in discretization of the
gradient term of $\phi^4$ theory we obtain a set of the overlap
integrals $J_{x}$, which can be chosen rather arbitrary.
Then an interesting question arises: if we can retain a lattice
in the bare theory, then what lattice model should be chosen?

The answer can be found from Eq.\,75. Since dependence on
$m/\Lambda$ is absent, we can tend this ratio to zero. But
in this limit (when $\xi/a\to \infty$) there are physical
grounds for independence of functions $F_i$ on
the way of cut-off. If such independence takes place for
$m/\Lambda\to 0$, it retains for arbitrary $m/\Lambda$ due to
independence of functions $F_i$ on this parameter. In fact, this
argumentation implies renormalizability of theory (due to which
the dependence on $\Lambda$ can be excluded) and belonging of the
lattice model to the proper universality class (inside of which
the dependence on the way of cut-off is absent).

The lattice theory is frequently considered as a reasonable
approximation to the true field theory.  In this case we should
accept the condition $ \xi \gg a$, which signifies that one has a
lot of lattice sites on the characteristic scale of variation of
field. This condition can be strengthen till $\xi/a\to \infty$ or
liberalized till $ \xi \agt a$. The first case corresponds to  the
point of phase transition and  gives $g=0$. In the second case we
obtain restriction $g\alt 1$ (for the proper charge normalization
 \cite{117}), which can be used to obtain the upper bound on
the Higgs mass \cite{617,701}.

In fact,  the lattice theory should not be considered as any
approximation to field theory, though it is possible for $g_0\ll
1$. The true field theory is continuous from the very beginning
and does not contain any lattice. The lattice is present only in
the bare theory, which is an auxiliary construction and is
completely removed later. No physical requirements, like $\xi\gg
a$,  are relevant for it.  If one removes the condition $\xi\gg
a$, then any values of $g$ become admissible\,\footnote{\,This
point of view is in complete agreement with mathematical
definitions  \cite{106,107}, according to which the limit $a\to 0$
is taken for the arbitrarily chosen dependencies $g_0(a)$ and
$m_0(a)$ (see Footnote 14). We impose conditions $g_0\to\infty$,
$g_0^{-1/2} m_0^2 a^2 \to -\infty$, $g_0^{-1} m_0^2 a^2 =-\kappa$,
necessary for transformation to the Ising model (Sec.\,6).  }. In
fact, a real designation of the bare theory is to represent the
relations between physical quantities in the parametric form (74).
Such representation has no deep sense already due to its
ambiguity: it can be written in many different forms, changing
$g_0$ and $m_0/\Lambda$ by any other pair of variables.

\vspace{2mm}

We see that contradiction between the continual and lattice
approaches is resolved by a special character of
renormalizability in $\phi^4$ theory: correct relations (77)
between physical quantities can be obtained for the arbitrary
value of the  parameter $a/\xi$, while a dependence on this
parameter is absent; to obtain the continuous renormalized
theory, there is no need to eliminate a lattice from the bare
theory.

\vspace{3mm}
\begin{center}
{\bf 9. General situation in renormalizable theories}
\end{center}

The interesting question arises: is such kind of
renormalizability related with the specific properties of
$\phi^4$ theory,  or it is a manifestation of some general
mechanism?

We shall see below that the second variant is correct. It can be
understood in the framework of Wilson's many-parameter
renormalization group (RG)  \cite{105}.
According to it, the parameters $p_i$
of some lattice Hamiltonian are considered as
functions of the length scale
${\it l}$.\,\footnote{\,Physically it is explained by the well-known
Kadanoff construction. In the description of magnetics,
one begins with the microscopic Hamiltonian for elementary
spins in the lattice sites. Then it is possible to introduce
the macroscopic spin variables corresponding to the blocks of size
${\it l}$ and  write the effective exchange Hamiltonian for
them.  Since the blocks of size  $n{\it l}$ can be composed of
$n^d$ blocks of size ${\it l}$, then recalculation
$p_i({\it l}) \to p_i(n{\it l})$ is possible, i.e.
$p_i(n{\it l})=H_i\left(n,\left\{p_k({\it l})\right\}\right) $.
Taking $n$ close to unity, one can obtain Eqs.\,78.}
The flow of
these parameters is determined by the RG equations, which can be
written in the differential form
$$
-\frac{d p_i}{d \ln ({\it l}/a)} = F_i\{p_k \} \,.
\eqno(78)
$$
These equations can be linearized near the fixed point
$$
p_i({\it l})=p^*_i \qquad \mbox{ (\,for\quad all\quad  {\it l\,})}
\eqno(79)
$$
and investigated by the standard methods of
linear algebra.  The ordinary phase transitions are described by
the saddle points of such equations. The simplest saddle point in
the two-parameter space has the straight-line trajectories in
two main directions (one stable and one unstable), while the rest
of trajectories are hyperbolic. For the usual phase transitions,
there are infinite number of stable directions and
one (in the simplest case)  unstable direction. The latter is
related with some controlling parameter like temperature,
measuring the distance to the critical point.

Instead of increasing  ${\it l}$ for fixed $a$, we can
diminish $a$ for fixed ${\it l}$.
The continuum limit  $a\to 0$
of field theory corresponds to
the critical surface $\xi/a=\infty$  in the
many-parameter space (Fig.14).
\begin{figure}
\centerline{\includegraphics[width=5.5 in]{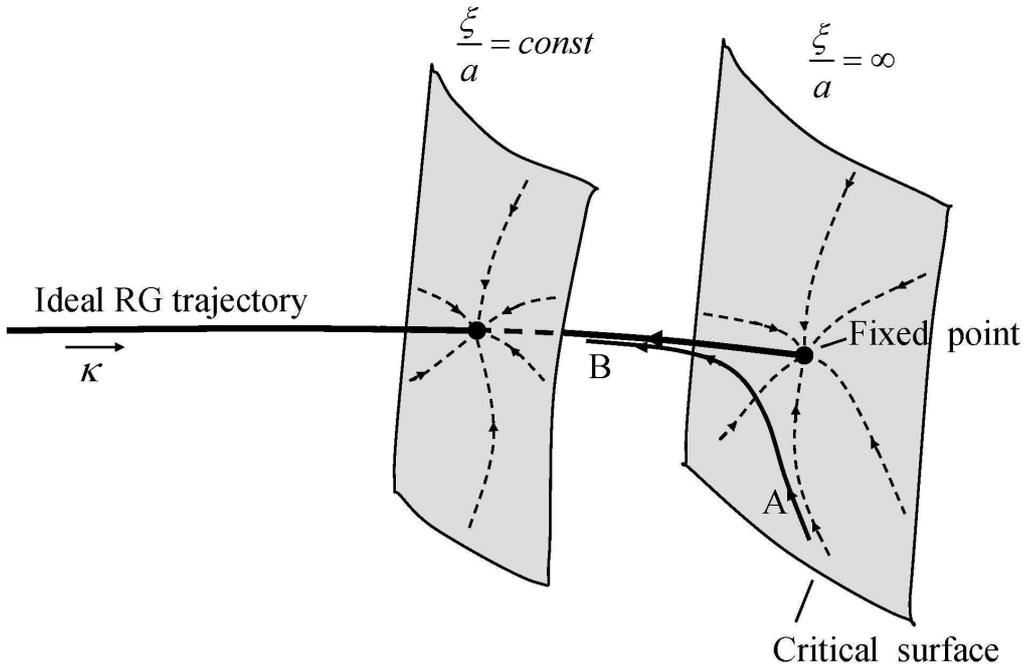}}
\caption{Schematic of the Wilson many-parameter space.}
\label{fig14}
\end{figure}
All trajectories on the critical
surface tend to the fixed point. The unstable trajectory,
originating at the fixed point will be referred as
an "ideal RG trajectory": along it one has the exact one-parameter
scaling, which is a pipe dream in many fields of physics
(see e.g. \cite{702}).
 To define it
rigorously, let consider the limit $a\to 0$ with fixed
$\xi/a$;  then all trajectories lying at the surface
$\xi/a=const$ (Fig.14) tend to one point (analogously
to the critical surface), while
the locus of such points is the ideal RG trajectory.

Let the parameter $\kappa$ is  measuring the
distance along the
ideal trajectory: then $\xi/a$
(or $\Lambda/m$) is a function
of $\kappa$. Analogously, all dimensionless quantities
depend only on $\kappa$, while the dimensional quantities
are measured in units of $\Lambda$. As a result, we come
to equations
$$
g=F_g\left( \kappa\right)\,, \qquad
m=\Lambda F_m\left( \kappa\right) \,, \qquad
A_i=\Lambda^{d_i} F_i\left( \kappa\right)\,,
\eqno(80)
$$
which coincide with (76) and give the relations (77)
without dependence on $m/\Lambda$.

The above construction has a following sense. If the limit
$a\to 0$ is taken in the arbitrary manner, then the
system will go to infinity along the unstable
direction and appear far from the critical surface,
which is our goal. Therefore, we suggest to take the continual
limit in two  steps:

(a) take a limit  $a\to 0$ for $a/\xi=const$;

(b) take a limit  $a/\xi\to 0$.

\noindent
It appears, that  dependence on $a$ in relations (77)
disappears already at the first step.
The second step becomes
unnecessary and there is no need to take the continuum limit in
the bare theory\,\footnote{\,These ideas are close to the QCD
specialists, and in fact the above consideration was partially
taken from "Introduction to lattice QCD" by R.\,Gupta \cite{703}.
This picture is discussed there in relation to improvement of the
lattice action, and the author claims that "simulations, done
along the ideal RG trajectory, will reproduce the continuum
physics without discretization errors". It implies the absence of
$a/\xi$ dependence, in accordance with our results.  Only final
conclusion was not made, that the continuum limit is not
necessary in the bare theory.}.
The $const$ appearing in (a) is one of the possible
definitions of the parameter $\kappa$.

Any RG trajectory is a line of "constant physics", since
the RG transformation  is simply a mental construction
and does not affect the large-scale properties of the system.
All trajectories belonging to the critical surface and
terminated in the fixed point, give the equivalent lattice
representations for the unique continuous field theory; another
equivalent representation is given by the ideal RG trajectory
originating from the fixed point.
Consider the trajectory $AB$, which begins
near the critical surface and goes along it, and then
tends to the ideal RG trajectory (Fig.14).
Introducing  $\tilde\kappa$ as a distance along $AB$,
we come to the parametric representation analogous to (80)
and relations (77), following from it. The latter relations
will be the same as obtained from (80), since in both cases
they are independent of $\xi/a$ and correspond to the physically
equivalent models for $\xi/a=0$ and $\xi/a=\infty$.  We can retain
definition of  $\kappa$ as a distance along the ideal trajectory,
and assign it to the point of $AB$, corresponding to the same
value of $\xi/a$.  Writing relations analogous to (80)
$$
g=\tilde F_g\left( \kappa\right)\,,\qquad
m=\Lambda F_m\left( \kappa\right) \,,\qquad
A_i=\Lambda^{d_i} \tilde F_i\left(\kappa\right)\,,
\eqno(81)
$$
we see that the second relation remained unchanged,
but the rest of them become different.
The charge  $g$ usually belongs  to irrelevant
parameters and we can introduce "the axis of charges" on
the critical surface; using trajectories of $AB$ type with
different directions relative to "the axis of charges"
we can obtain different functions $g=\tilde F_g(\kappa)$.
It means that the functional relation between $g$ and
$\kappa$ becomes indeterminate and can be omitted.

As a result, the renormalized and bare sectors of theory become
decoupled. The renormalized sector contains relations
(77), where  $g$ and  $m$ are considered as independent variables.
The bare sector contains only relation
$a/\xi=m/\Lambda=F_m\left(\kappa\right)$, which determines
$\kappa$ as a function of $a$ and is irrelevant  from
viewpoint of  physics. Parameter  $a/\xi$ becomes
absolutely free.

Thereby, we come to the following conclusion:
renormalizable theory of the considered type allows
representation in the form of  lattice theory, which gives
the correct relations between physical quantities, and contains
free parameter $a/\xi$, which does not enter these relations.

\vspace{3mm}
\begin{center}
{\bf 10. Application to theory of confinement} \end{center}

Consider a pure Yang-Mills theory,  i.e. QCD without quarks. Then
the quark mass does not enter as a parameter and the theory
contains no natural mass scale.  To avoid the specific
difficulties related with such situation, we introduce the
"extended version" of Yang-Mills theory, where the role of the
bare mass $m_0$ (more exactly, the ratio $m_0/\Lambda$) is played
by some auxiliary parameter $p$ characterizing the lattice theory;
as a renormalized mass, we accept the mass  $m$ of the lightest
glueball (the bound state of several gluons), while the
correlation length $\xi$ is defined as $m^{-1}$. Thereby,  two
bare parameters  $g_0$ and $p$ provide the observable values for
renormalized $g$ and $m$.  In order to return to the standard
variant of theory, we should remove the introduced extra degree of
freedom by fixing one relation between observable quantities.
However, it can be done on the late stage (see below), while the
main of analysis is produced for the "extended version". The
latter is analogous to $\phi^4$ theory.

According to Wilson \cite{704}, confinement can be proved in the
lattice version of the Yang-Mills theory for large values of the
bare charge $g_0$. The energy of interaction for two probe
quarks separated by a distance $R$ is $V(R)=\sigma R$, while the
string tension $\sigma$ and the glueball mass $m$ are given by
expressions  \cite{703,704,705}
$$
\sigma= \frac{\ln (3 g_0^2)}{a^2}
\,, \qquad m = \frac{4\,\ln (3 g_0^2)}{ a}
\,.
\eqno(82)
$$
In spite of the evident success, the Wilson theory is
considered as purely illustrative and having no relation
to real QCD. As was indicated by Wilson himself, his
theory corresponds to a situation
$$
\xi\ll a  \qquad \mbox{ or}\qquad  m\gg \Lambda \,,
\eqno(83)
$$
which is considered as nonphysical. An attempt to advance into the
physical region inevitably destroys the strong coupling regime.
Indeed, the $\beta$-function in terms of the bare charge is
believed to be negative \cite{706,707}, and $g_0$ tends to zero in
the continuum limit $a\to 0$. Therefore, the strong coupling
regime is inevitably destroyed in the physical region $\xi/a\agt
1$ and Wilson's theory becomes inapplicable.

\vspace{2mm}

A situation changes drastically, if we use representation
(76,\,77) introduced in the previous sections. In this case:

(i) due to absence of the $\xi/a$ dependence in (77), this
parameter can be taken arbitrary: it eliminates objections against
the nonphysical regime in Wilson's theory.

(ii) there is no direct relation between the bare and renormalized
charge; rewriting the second expression (82) in the
form
$$
g_0^2=\frac{1}{3} \exp\left( \frac{m a}{4}\right) =
      \frac{1}{3} \exp\left\{ \frac{a}{4\xi} \right\}\,,
\eqno(84)
$$
we see that, independently of renormalized values of
$g$ and $m$, it is possible to  choose the free parameter $a/\xi$
so as to  obtain a sufficiently large value for $g_0$.  Then
Wilson's theory becomes applicable and the first relation (82)
gives  a finite value for  $\sigma$, i.e.
confinement.

Representation (76,\,77) cannot be introduced for the simplest
Wilson action \cite{703,704,705}, since it does not contain a
sufficient number of parameters.  To obtain the observable
values of $\sigma$ and $m$ one should fix both $g_0$ and $a$; but
the fixed $a$ means  impossibility to introduce  a
representation with free parameter $a/\xi$. This problem can be
easily solved using more complicated forms of the action
\cite{703}. As a result, we obtain instead of (82)
$$
\sigma= \frac{\ln (3 g_0^2)}{a_1^2}
\,, \qquad m = \frac{4\,\ln (3 g_0^2)}{ a_2} \,,
\eqno(85)
$$
where  $a_1=k_1 a$, $a_2=k_2 a$ simply by dimensional reasons
\cite{118}.

Relation (77) in application to $\sigma$ has a form
$$
\sigma=m^2 F_\sigma(g)\,,
\eqno(86)
$$
so $g$ is functionally related with $\sigma/m^2$. On the
other hand, Eqs.\,85 give
$$
\frac{\sigma}{m^2} = \frac{a_2^2}{ 16 a_1^2  \,\ln (3 g_0^2)} \,,
\eqno(87)
$$
and the ratio $\sigma/m^2\,$ changes from a finite value till zero
in the strong coupling region $g_0\agt 1$. It means that only
restricted range of $g$ values can be reproduced.  Such
restriction is natural due to the physical essence of the problem.
Indeed,  the linear confinement potential is expected only at
large distances, where  $g$ is certainly not small; hence, small
values of  $g$ are inaccessible in the Wilson regime.  Contrary,
the restricted range of  $\sigma/m^2$ values goes across the logic
of theory. Indeed,  $a/\xi$ is a free parameter and all physical
results can be obtained at its arbitrary value. In the case
$a/\xi\gg 1$, the regime of confinement is controlled analytically
and any physically accessible value of  $\sigma/m^2$ should be
possible in this limit. In fact, the range of $\sigma/m^2$ values
can be extended if we use the models with essentially different
$a_1$ and $a_2$ \cite{118}. Absence of restrictions on
$\sigma/m^2$ in the presence of restrictions on  $g$ is possible
only if dependence $\sigma/m^2= F_\sigma(g)$ is singular (Fig.15);
fortunately, we can demonstrate that it is a probable variant.

Investigations of  complicated lattice versions of
Yang-Mills theory \cite{703} show existence of phase
transitions (lying in the region  $g_0\sim 1$), corresponding
to vanishing of the lightest glueball mass $m$, with finite
values  of  $\sigma$ and other mass parameters. These transitions
are considered as lattice artifacts, since they do not survive
in the continuum limit, when  $g_0\to 0$.  In our
approach the limit  $a\to 0$ is not necessary and such phase
transitions acquire the physical sense.  If $m$  vanishes at
the point $g=g^*$, then dependence $\sigma/m^2= F_\sigma(g)$
has a form shown in Fig.15.

\begin{figure}
\centerline{\includegraphics[width=4.1 in]{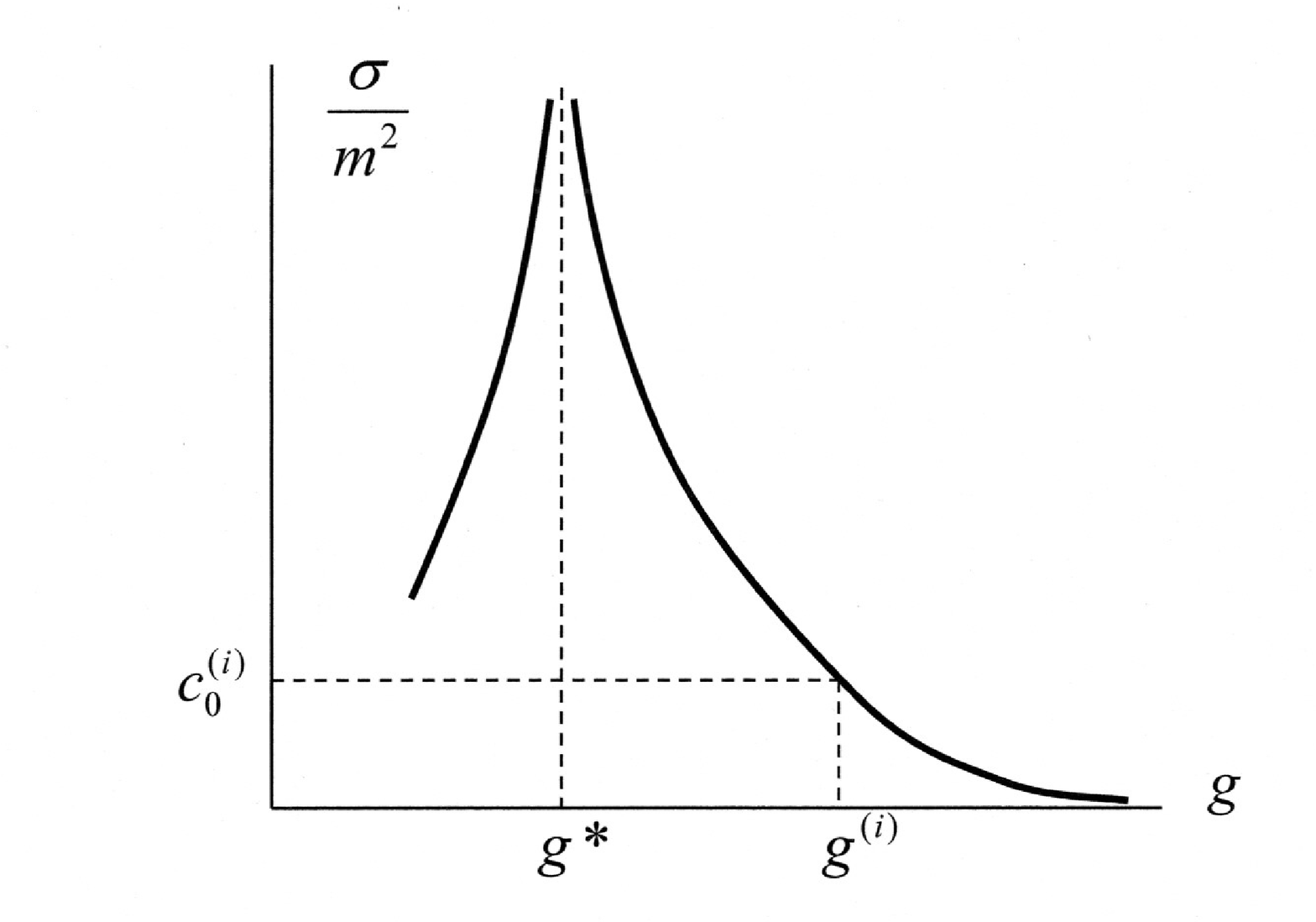}}
\caption{Dependence of $\sigma/m^2$ against $g$. In order to
obtain the special values $c^{(i)}_0$, corresponding to zero
values of the mass gap, one should mark all stable fixed points
$g^{(i)}$  on the horizontal axis and make a construction shown in
the figure.} \label{fig15}
\end{figure}

Existence of points
with $m=0$ in the parametric space means that
the "extended version" of Yang-Mills theory does not possess the
mass gap. To eliminate this defect, we should return to the
standard variant of theory, fixing one relation between  the
observable quantities. The character  of such relations
is well-known and is determined by  "dimensional
transmutation"  \cite[Sec.\,14.1]{703},
\cite[Sec.\,IV.6]{708}, according to which all quantities of the
same dimensionality differ only by the constant factor,
independent of  $g_0$. For our purposes
it is convenient to accept the condition
$$
\sigma/m^2=c\,,
\eqno(88)
$$
which defines the one-parameter family of Yang-Mills
theories with different values of the structural constant
$c$. Under condition  (88), the points with  $m=0$,
$\sigma=const$ become inaccessible.

It  does not yet prove the existence of a mass gap,
since $\sigma$ and $m$  can  vanish simultaneously.
In order to analyze such situations,
consider  the Gell-Mann -- Low equation for
the renormalized charge $g$ attributed to the length scale
$m^{-1}$
$$
\frac{d\, g^2}{d \,\ln m^2} =\beta(g^2)=
          \beta_0 g^4+ \beta_1 g^6+\ldots   \,,
\eqno(89)
$$
where  $\beta$-function does not coincide with that of the bare
theory, but has the same first coefficients  $\beta_0$ and
$\beta_1$. It is clear that the value  $g^*$ (Fig.15) is a root of
the $\beta$-function; generally, it has several roots determining
the RG fixed points. In the limit $m\to 0$, the charge $g$ tends
to one of these fixed points, while following variants are
possible for $\sigma/m^2$:  (a) $\sigma/m^2 \to \infty$, (b)
$\sigma/m^2 \to 0$, (c) $\sigma/m^2 \to c_0$.  The first two
variants are incompatible with Eq.\,88, while the third variant is
possible in the case  $c=c_0$. If there are several stable fixed
points $g^{(i)}$, then there are several special values
$c^{(i)}_0$ (see Fig.15), for which the mass gap vanishes; for
all other values of  $c$ the mass gap is finite.

Physically, it looks most probable that only one fixed point
$g^*$ with $\sigma/m^2 \to \infty$ is present, so no special
values $c^{(i)}_0$ arise. Mathematically, one can suggest an
infinite number of fixed points, which form a sequence
$c^{(i)}_0$ everywhere dense in the interval $(0,\infty)$.
However, small values of  $\sigma/m^2$ correspond to
the Wilson regime where finiteness of  $\sigma$ and $m$ is
verified
%straightforwardly.
immediately.
%directly.
As a result, the proof of the
mass gap is complete for small values of the structural
constant  $c$.

If quarks with a zero  mass\,\footnote{\,In the case of fermions,
the renormalization of mass has a multiplicative character and
the choice of the zero bare mass provides the zero renormalized
mass.  } are introduced, then the regime of dimensional
transmutation is conserved and the trick with "extension" of
theory remains possible; it seems, that the general structure of
theory is also retained.

\vspace{2mm}

In conclusion, the properties of continuous  Yang-Mills theory
can be reproduced by a certain lattice theory. The bare
charge $g_0$ in this lattice theory can be taken arbitrary,
and in particular infinitely large. For large $g_0$, any
reasonable lattice version of Yang-Mills theory gives finite
values of $\sigma$ and $m$. Vanishing of $m$  is possible
under exceptional conditions, which are
avoided in the general situation.  As a result, the problem of
analytical proof of confinement and the mass gap can be
 considered as solved, at least on the physical level of rigor.

\begin{center}
{\bf 11. Conclusion}
\end{center}

We have given an overview of field theory evolution from its
early stage to the present time. A widespread opinion on
triviality of $\phi^4$ theory and QED should in fact be
revisited. They are trivial in Wilson's sense, but escape
the Landau "zero-charge" situation. This conclusion is reached by
summation of weak-coupling perturbation expansions for
$\beta(g)$, analytical calculation of its strong coupling
asymptotics, and confirmation of results by analysis of strong
 coupling expansions. The second possibility in
 the Bogoliubov and Shirkov classification is shown to
 be actual, allowing to  construct  the continuous theory
 with finite interaction at large distances. A possibility to
 retain a lattice in the bare theory allows to justify the Wilson
 approach in theory of confinement.

\end{document}